\documentclass[twocolumn]{aastex631}

\usepackage{comment}

\begin{document}

\title{exoALMA XXII: A Two-dimensional Atlas of Deviations from Keplerian Disks}

\author[0000-0003-1117-9213]{Misato Fukagawa}
\affiliation{Astronomical Institute, Graduate School of Science, Tohoku University, 6-3 Aoba, Aramaki, Aoba-ku, Sendai, Miyagi 980-8578 Japan}
\affiliation{National Astronomical Observatory of Japan, 2-21-1 Osawa, Mitaka, Tokyo 181-8588, Japan}
\email{misato.fukagawa@astr.tohoku.ac.jp}

\author[0000-0001-8446-3026]{Andr\'es F. Izquierdo} 
\affiliation{Department of Astronomy, University of Florida, Gainesville, FL 32611, USA}
\affiliation{Leiden Observatory, Leiden University, P.O. Box 9513, NL-2300 RA Leiden, The Netherlands}
\affiliation{NASA Hubble Fellowship Program Sagan Fellow}

\author[0000-0002-0491-143X]{Jochen Stadler} 
\affiliation{Universit\'e C\^ote d'Azur, Observatoire de la C\^ote d'Azur, CNRS, Laboratoire Lagrange, 06304 Nice, France}
\affiliation{European Southern Observatory, Karl-Schwarzschild-Str. 2, D-85748 Garching bei M\"unchen, Germany}

\author[0000-0002-7212-2416]{Lisa W\"olfer} 
\affiliation{Department of Earth, Atmospheric, and Planetary Sciences, Massachusetts Institute of Technology, Cambridge, MA 02139, USA}

\author[0000-0002-5503-5476]{Maria Galloway-Sprietsma}
\affiliation{Department of Astronomy, University of Florida, Gainesville, FL 32611, USA}

\author[0000-0003-4039-8933]{Ryuta Orihara} 
\affiliation{Department of Astronomy, Graduate School of Science, The University of Tokyo, Tokyo 113-0033, Japan}

\author[0000-0001-8877-4497]{Masataka Aizawa} 
\affiliation{College of Science, Ibaraki University, 2-1-1 Bunkyo, Mito, Ibaraki 310-8512, Japan}

\author[0000-0002-3001-0897]{Munetake Momose} 
\affiliation{College of Science, Ibaraki University, 2-1-1 Bunkyo, Mito, Ibaraki 310-8512, Japan}
\affiliation{NAOJ Chile, National Astronomical Observatory of Japan, Los Abedules 3085 Oficina 701, Vitacura 7630414, Santiago, Chile}

\author[0000-0003-4679-4072]{Daniele Fasano} 
\affiliation{Universit\'{e} C\^{o}te d'Azur, Observatoire de la C\^{o}te d'Azur, CNRS, Laboratoire Lagrange, France}

\author[0000-0002-7695-7605]{Myriam Benisty}
\affiliation{Max-Planck Institute for Astronomy (MPIA), Königstuhl 17, 69117 Heidelberg, Germany}

\author[0000-0003-1534-5186]{Richard Teague}
\affiliation{Department of Earth, Atmospheric, and Planetary Sciences, Massachusetts Institute of Technology, Cambridge, MA 02139, USA}

\author[0000-0003-4689-2684]{Stefano Facchini}
\affiliation{Dipartimento di Fisica, Universit\`a degli Studi di Milano, Via Celoria 16, 20133 Milano, Italy}

\author[0000-0001-5907-5179]{Christophe Pinte}
\affiliation{Univ. Grenoble Alpes, CNRS, IPAG, 38000 Grenoble, France}
\affiliation{Monash Centre for Astrophysics (MoCA) and School of Physics and Astronomy, Monash University, Clayton Vic 3800, Australia}

\author[0000-0003-2253-2270]{Sean M. Andrews}
\affiliation{Center for Astrophysics | Harvard \& Smithsonian, Cambridge, MA 02138, USA}

\author[0000-0001-7258-770X]{Jaehan Bae}
\affiliation{Department of Astronomy, University of Florida, Gainesville, FL 32611, USA}

\author[0000-0001-6378-7873]{Marcelo Barraza-Alfaro}
\affiliation{Department of Earth, Atmospheric, and Planetary Sciences, Massachusetts Institute of Technology, Cambridge, MA 02139, USA}

\author[0000-0002-2700-9676]{Gianni Cataldi} 
\affiliation{National Astronomical Observatory of Japan, 2-21-1 Osawa, Mitaka, Tokyo 181-8588, Japan}

\author[0000-0003-2045-2154]{Pietro Curone} 
\affiliation{Departamento de Astronomía, Universidad de Chile, Camino El Observatorio 1515, Las Condes, Santiago, Chile}

\author[0000-0002-1483-8811]{Ian Czekala}
\affiliation{School of Physics \& Astronomy, University of St. Andrews, North Haugh, St. Andrews KY16 9SS, UK}

\author[0000-0002-9298-3029]{Mario Flock} 
\affiliation{Max-Planck Institute for Astronomy (MPIA), Königstuhl 17, 69117 Heidelberg, Germany}

\author[0000-0002-5910-4598]{Himanshi Garg}
\affiliation{School of Physics and Astronomy, Monash University, Clayton VIC 3800, Australia}

\author[0000-0002-8138-0425]{Cassandra Hall} 
\affiliation{Department of Physics and Astronomy, The University of Georgia, Athens, GA 30602, USA}
\affiliation{Center for Simulational Physics, The University of Georgia, Athens, GA 30602, USA}
\affiliation{Institute for Artificial Intelligence, The University of Georgia, Athens, GA, 30602, USA}

\author[0000-0001-6947-6072]{Jane Huang} 
\affiliation{Department of Astronomy, Columbia University, 538 W. 120th Street, Pupin Hall, New York, NY, USA}

\author[0000-0003-1008-1142]{John~D.~Ilee} 
\affiliation{School of Physics and Astronomy, University of Leeds, Leeds, UK, LS2 9JT}

\author[0009-0007-5371-3548]{Jensen Lawrence}
\affiliation{Department of Earth, Atmospheric, and Planetary Sciences, Massachusetts Institute of Technology, Cambridge, MA 02139, USA}

\author[0000-0002-8896-9435]{Geoffroy Lesur} 
\affiliation{Univ. Grenoble Alpes, CNRS, IPAG, 38000 Grenoble, France}

\author[0000-0002-2357-7692]{Giuseppe Lodato} 
\affiliation{Dipartimento di Fisica, Universit\`a degli Studi di Milano, Via Celoria 16, 20133 Milano, Italy}

\author[0000-0003-4663-0318]{Cristiano Longarini} 
\affiliation{Institute of Astronomy, University of Cambridge, Madingley Road, CB3 0HA, Cambridge, UK}
\affiliation{Dipartimento di Fisica, Universit\`a degli Studi di Milano, Via Celoria 16, 20133 Milano, Italy}

\author[0000-0002-8932-1219]{Ryan A. Loomis}
\affiliation{National Radio Astronomy Observatory, 520 Edgemont Rd., Charlottesville, VA 22903, USA}

\author[0000-0002-1637-7393]{Fran\c{c}ois M\'enard}
\affiliation{Univ. Grenoble Alpes, CNRS, IPAG, 38000 Grenoble, France}

\author[0000-0002-4716-4235]{Daniel J. Price} 
\affiliation{School of Physics and Astronomy, Monash University, Clayton VIC 3800, Australia}

\author[0000-0003-4853-5736]{Giovanni Rosotti} 
\affiliation{Dipartimento di Fisica, Universit\`a degli Studi di Milano, Via Celoria 16, 20133 Milano, Italy}

\author[0000-0003-1412-893X]{Hsi-Wei Yen} 
\affiliation{Academia Sinica Institute of Astronomy \& Astrophysics, 11F of Astronomy-Mathematics Building, AS/NTU, No.1, Sec. 4, Roosevelt Rd, Taipei 10617, Taiwan}

\author[0000-0001-8002-8473]{Tomohiro C. Yoshida} 
\affiliation{National Astronomical Observatory of Japan, 2-21-1 Osawa, Mitaka, Tokyo 181-8588, Japan}
\affiliation{Department of Astronomical Science, The Graduate University for Advanced Studies, SOKENDAI, 2-21-1 Osawa, Mitaka, Tokyo 181-8588, Japan}

\author[0000-0002-3468-9577]{Gaylor Wafflard-Fernandez} 
\affiliation{Univ. Grenoble Alpes, CNRS, IPAG, 38000 Grenoble, France}

\author[0000-0003-1526-7587]{David J. Wilner} 
\affiliation{Center for Astrophysics | Harvard \& Smithsonian, Cambridge, MA 02138, USA}

\author[0000-0002-7501-9801]{Andrew J. Winter}
\affiliation{Astronomy Unit, School of Physics and Astronomy, Queen Mary University of London, London E1 4NS, UK}

\author[0000-0001-9319-1296]{Brianna Zawadzki} 
\affiliation{Department of Astronomy, Van Vleck Observatory, Wesleyan University, 96 Foss Hill Drive, Middletown, CT 06459, USA}
\affiliation{Department of Astronomy \& Astrophysics, 525 Davey Laboratory, The Pennsylvania State University, University Park, PA 16802, USA}

\begin{abstract}
Protoplanetary disks are the birthplaces of planetary systems, and deviations from Keplerian rotation imprinted in disk gas kinematics serve as key tracers of physical processes and the presence of protoplanets within disks. 
Using the the CO ($J$=3--2) data from the exoALMA Large Program encompassing 15 disks, we constructed two-dimensional (2D) maps of centroid velocity, line width, and peak intensity, and extracted non-Keplerian deviations by subtracting smooth Keplerian models.
This paper provides the first systematic and uniform overview of 2D gas substructures across the entire exoALMA sample.
We find that all targets exhibit large-scale deviations from smooth Keplerian disks, displaying a variety of morphologies including spiral-like structures, arc- or ring-like features, and patterns indicative of variations in the emitting surface height. Non-axisymmetric spiral-arm features are detected or suggested in five disks (CQ Tau, MWC 758, HD 135344B, HD 34282, and SY Cha), and are preferentially found in Herbig Ae/Fe systems. In contrast, some other sources (J1852, PDS 66, and V4046 Sgr), despite exhibiting noticeable deviations, appear to be dynamically quieter. This 2D atlas suggests that kinematic substructures are ubiquitous in large ($\gtrsim 100$~au) protoplanetary disks with ages of a few million years, based on the observations obtained with sufficient sensitivity at moderate-to-high spatial resolution of $\sim$20~au and high velocity resolution of $\sim$0.1 km~s$^{-1}$.
\end{abstract}


\section{Introduction} \label{sec:intro}
The dynamical structure of protoplanetary disks not only indirectly indicates the presence of planets embedded within the disk \citep{Pinte2023PPVII}, but also reflects the underlying fundamental disk properties, such as density and temperature distributions \citep[e.g.,][]{Teague2018_as209}, as well as the mechanisms driving the disk dynamics \citep[e.g.,][]{Speedie2024}. The Large Program exoALMA with the Atacama Large Millimeter/submillimeter Array (ALMA) aims to provide such fundamental insights into planetary-system formation by observing and analyzing 15 protoplanetary disks with unprecedentedly high velocity resolution and correspondingly high spatial resolution. The first series of exoALMA papers reported a wide range of results mainly from CO gas data analyses, such as measurements of pressure gradients, temperature, rotation curves, surface density distributions, as well as detailed structures of the dust continuum emission \citep[e.g.,][]{exoALMA_surface,exoALMA_pressure,exoALMA_rotation_curve,exoALMA_dust}. They also included detailed investigations of several disks that exhibit intriguing fine structures and spectral features \citep[e.g.,][]{exoALMA_asymmetries,exoALMA_lkca15,exoALMA_systematics,exoALMA_pressure_broadening}.

Creating two-dimensional (2D) images regarding the spatial distribution of line properties (moment maps) is almost always the first step in data analysis for gas observations. Recent developments in image-analysis techniques and tools have enabled the extraction of disk structures from high-fidelity ALMA images, allowing subtle deviations from Keplerian motion and complex substructures to be identified in several large and bright disks \citep[e.g.,][]{Teague2022,Teague2021,Izquierdo2023}. 
This paper presents the 2D moment images and other distilled representations of the data cubes (hereafter referred to as ``moment maps" for simplicity) of 15 protoplanetary disks obtained in exoALMA. We provide an overview of how the observed velocity and intensity distributions deviate from those expected for a Keplerian rotating disk, focusing on the centroid velocity of the spectral line, line width, and peak intensity derived through parametric disk-model fitting described in detail by \citet{exoALMA_discminer}.

The paper is organized as follows. A brief description on the observation program and data reduction is given in Section 2. The moment maps, together with their comparisons to a smooth Keplerian disk (residual images), are presented and explained in Section 3. A summary is provided in Section 4.

\section{Observations and Data Reduction} \label{sec:sample}
The full description of the exoALMA sources and observations are provided by \citet{exoALMA1}. The data were calibrated and imaged as detailed in \citet{exoALMA_calimg}. This paper uses the moment maps and the residual images obtained through disk modeling, following the analysis method presented in \citet{exoALMA_discminer}. The relevant points are briefly summarized below.
While a synthesized beam size of $0\farcs15$ and a velocity resolution of 100 m\,s$^{-1}$ are used as the baseline, the optimal combination is adopted during the disk modeling process, primarily to ensure a sufficient signal-to-noise ratio (S/N). Depending on the target and molecular species, images with beam sizes of $0\farcs15$ or $0\farcs30$ and velocity resolutions of 28 m\,s$^{-1}$, 100 m\,s$^{-1}$, or 200 m\,s$^{-1}$ are utilized (Table~\ref{tab:beam_chan} in Appendix).

The extraction of moment maps and the fitting of a smooth Keplerian disk to the data from the observations were performed using \textsc{discminer}\footnote{https://github.com/andizq/discminer} \citep{Izquierdo2021}.
It allows us to extract moment maps through the line profile fitting. Depending on the disk inclination and the optical thickness of the emission lines, radiation from the back-side emitting surface may be blended with that from the upper surface along the line of sight. Consequently, producing moment maps suitable for scientific interpretation is not straightforward. In the exoALMA analysis, single-component (Gaussian, bell) and double-component (double-bell, double-Gaussian) fitting kernels were all tested and the one that best reproduced the observed spectral profiles was selected for each source and each molecular species. 
With respect to the fitting of the Keplerian rotating disk model, \textsc{discminer} does not assume any particular physical process, but prepares a parametric model to be fitted to the observed channel maps. As such, it allows the extraction of disk dynamical information without being restricted by an assumed physical structure too much. The details of the modeling, along with the best-fit parameters, regarding the disk orientation, systemic velocity, emission surface, and line profile for individual source, are described in \citet{exoALMA_discminer}. 
The data underlying this study, including the generated line cubes and residual images, are made available at: https://www.exoalma.com/home 

Since the primary focus of this paper is on the substructures of the molecular gas with as much spatial details as possible, the data from $^{12}$CO ($J$=3--2), which have high S/N, are mainly used. The images in $^{13}$CO ($J$=3--2) and CS ($J$=7--6) are included in the Appendix~\ref{sec:appendix_13co} and \ref{sec:appendix_cs}. The velocity and intensity distributions of $^{13}$CO are similar to those of $^{12}$CO, and can therefore be used to support the reliability of the features seen in $^{12}$CO. In contrast, the CS emission often appears markedly different from CO, indicating that the CO-to-CS ratio varies significantly from source to source; its physical and chemical properties will be discussed in separate papers (G. Cataldi et al., in prep.).

The azimuthally averaged radial intensity and velocity profiles for the molecular emission lines are presented in \citet{exoALMA_surface} and \citet{exoALMA_pressure}. For global, symmetric substructures such as concentric rings and gaps with low structural contrast, we refer the reader to those and other exoALMA papers from the first series \citep[e.g.,][]{exoALMA_rotation_curve}, since these radial profiles allow us to identify symmetric substructures with a higher S/N compared to direct inspections on the 2D images. As another complementary study, the impact of non-Keplerian features on individual intensity channels is analyzed in \citet{exoALMA_kink}.

\section{Results} \label{sec:morphology}
\subsection{Moment Maps}
\begin{figure*}
   \centering
   \vspace{1em}   
   \includegraphics[width=\textwidth]{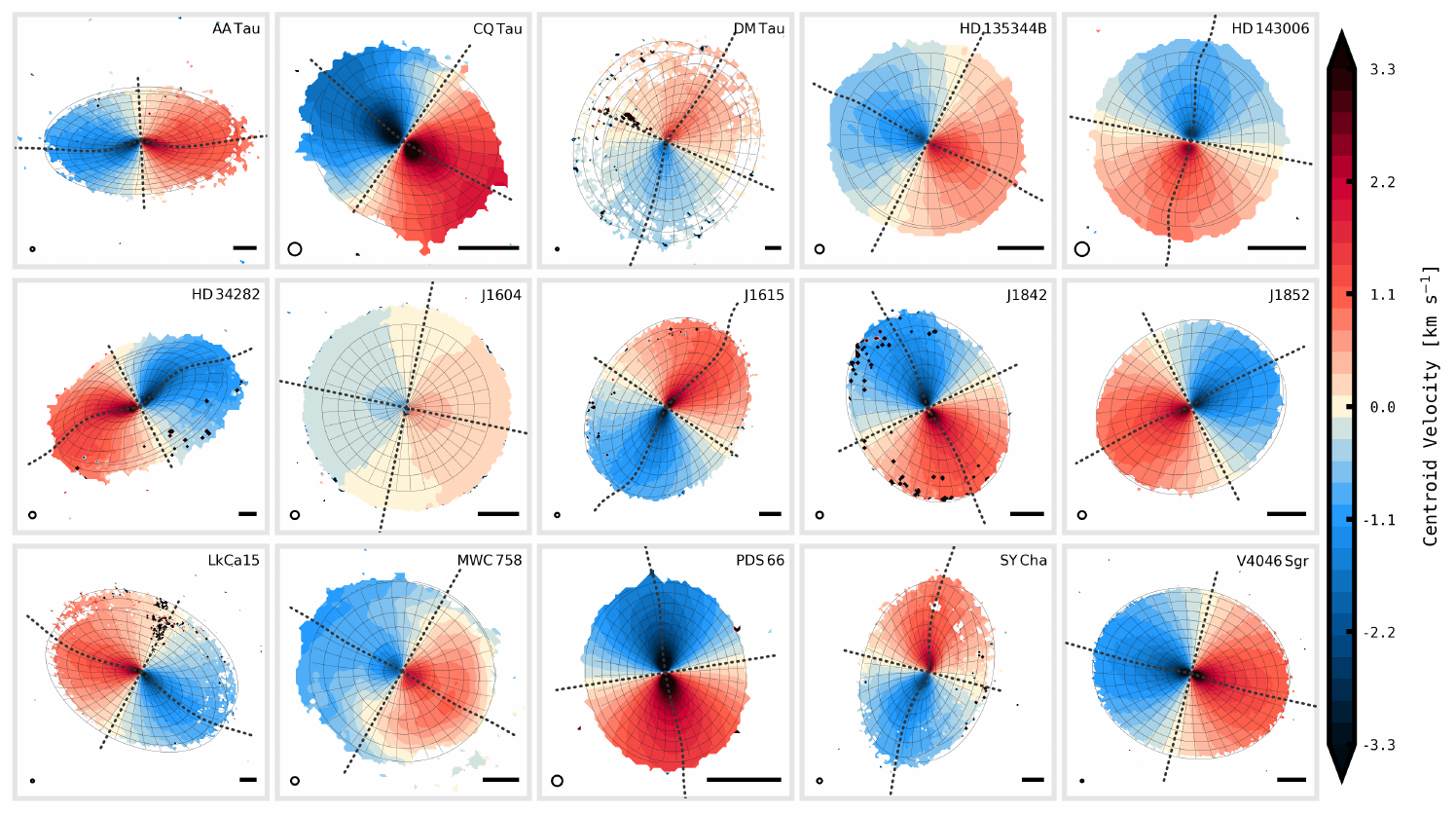}
    \caption{Velocity (line centroid) images in $^{12}$CO (3--2) obtained with \textsc{discminer}. All panels are shown using a common color scale, as indicated by the color bar on the right. The beam size is the same at $0\farcs15$ for all the sources and is indicated in the lower left corner of each panel, though the panel size varies. The black line shown in the lower right corner of each panel represents a spatial scale of 100 au. Pixels with S/N below 4.5 are masked. The major and minor axes of the fitted disk model are shown as black dashed lines. The emitting surfaces are overlaid as thin gray contours at radial intervals corresponding to integer multiples of the beam size (1--4 beams, depending on the panel), chosen to avoid overcrowding.}
    \label{fig:velocity_12co}
\end{figure*}

\begin{figure*}
   \centering
   \vspace{1em}   
   \includegraphics[width=\textwidth]{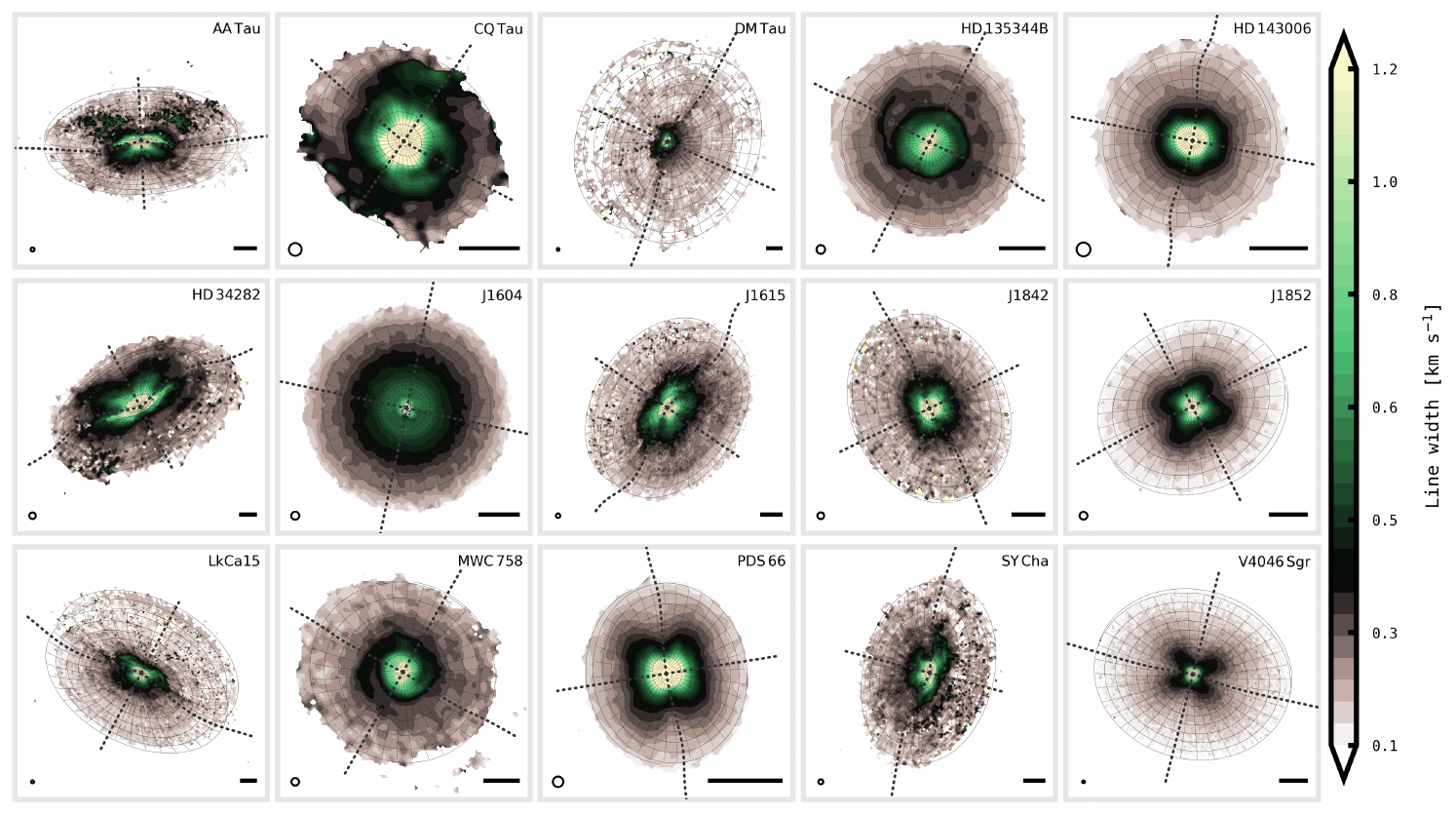}
    \caption{The same as Figure~\ref{fig:velocity_12co} but for the line width. Pixels with S/N below 4.5 are masked. }
    \label{fig:linewidth_12co}
\end{figure*}

The images of the centroid-velocity and line-width distributions in $^{12}$CO are shown in Figures~\ref{fig:velocity_12co} and \ref{fig:linewidth_12co}, respectively. The velocity distribution in each disk is broadly consistent with the Keplerian rotation on large scales. However, whereas a Keplerian disk should show a line-of-sight velocity pattern symmetric about the disk minor axis (corresponding to the systemic velocity, $v$=0, in Figure~\ref{fig:velocity_12co}), MWC~758 shows a lack of such symmetry, suggesting non-Keplerian motion. Although the amplitudes of the wiggles around the minor axis are small, CQ~Tau and HD~135344B also show hints of asymmetry. Moreover, Figure~\ref{fig:linewidth_12co} suggests that the line widths of these three objects (CQ~Tau, HD~135344B, and MWC~758) tend to be larger ($>$0.4 km s$^{-1}$) beyond $\sim$100~au, compared to those at the same radii in the other sources. 

\begin{figure*}
   \centering
   \vspace{1em}   
   \includegraphics[width=\textwidth]{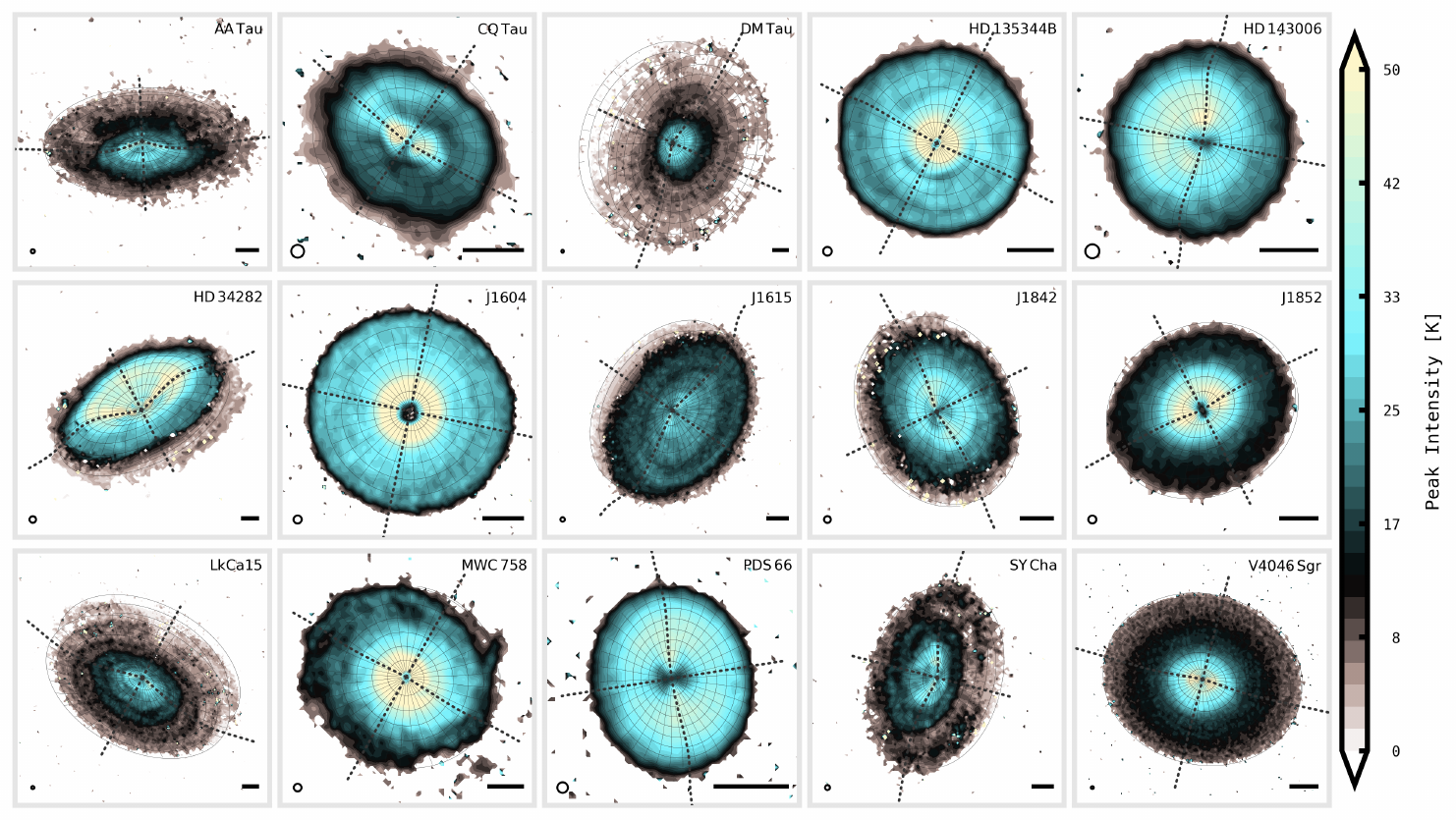}
    \caption{The same as Figure~\ref{fig:velocity_12co} but for the peak intensity. Pixels with S/N below 4 are masked. 
    }
    \label{fig:peakint_12co}
\end{figure*}

The peak intensity (moment 8) images in $^{12}$CO are shown in Figure~\ref{fig:peakint_12co}.
These images suggest non-axisymmetric features in the disks of CQ Tau and MWC~758, as well as annular or arm-like substructures in AA~Tau, DM~Tau, HD~135344B, HD~34282, J1615, J1842, LkCa~15, and SY Cha. These signs of substructures become more evident after subtracting the background disk structure, as presented in the next subsection.

The inclined disks of AA~Tau, HD~143006, HD~34282, J1615, and SY~Cha show pronounced vertical height. It is evident that they are vertically flared since, in such disks, when observing the upper emission surface from the center outward, the region of high peak intensity does not follow a straight line along the disk major axis but instead deviates upward, indicating an increasing height with radius \citep{Rosenfeld2013}.
In comparison, the degree of bending of the bright ridge along the major axis is smaller in $^{13}$CO, confirming that emission from a lower vertical layer is traced in $^{13}$CO than $^{12}$CO. This has been quantitatively measured by  \citet{exoALMA_surface}; they calculated the representative values of the height-to-radius ratio $\langle z/r \rangle$ for 13 disks and derived average values across multiple disks, obtaining $\langle$$z/r$$\rangle$=0.28 and 0.16  for $^{12}$CO and $^{13}$CO, respectively (excluding PDS~66, which yielded a ratio of zero). 

Due to the large optical thickness of the CO emission, the brightness temperature can be regarded as a reasonable indicator of the kinetic temperature of the molecular gas. The analysis of the temperature structures of the exoALMA sources were also reported in \citet{exoALMA_surface}.

\subsection{Residuals of the Centroid Velocity}
\begin{figure*}
    \centering
    \includegraphics[width=\textwidth]{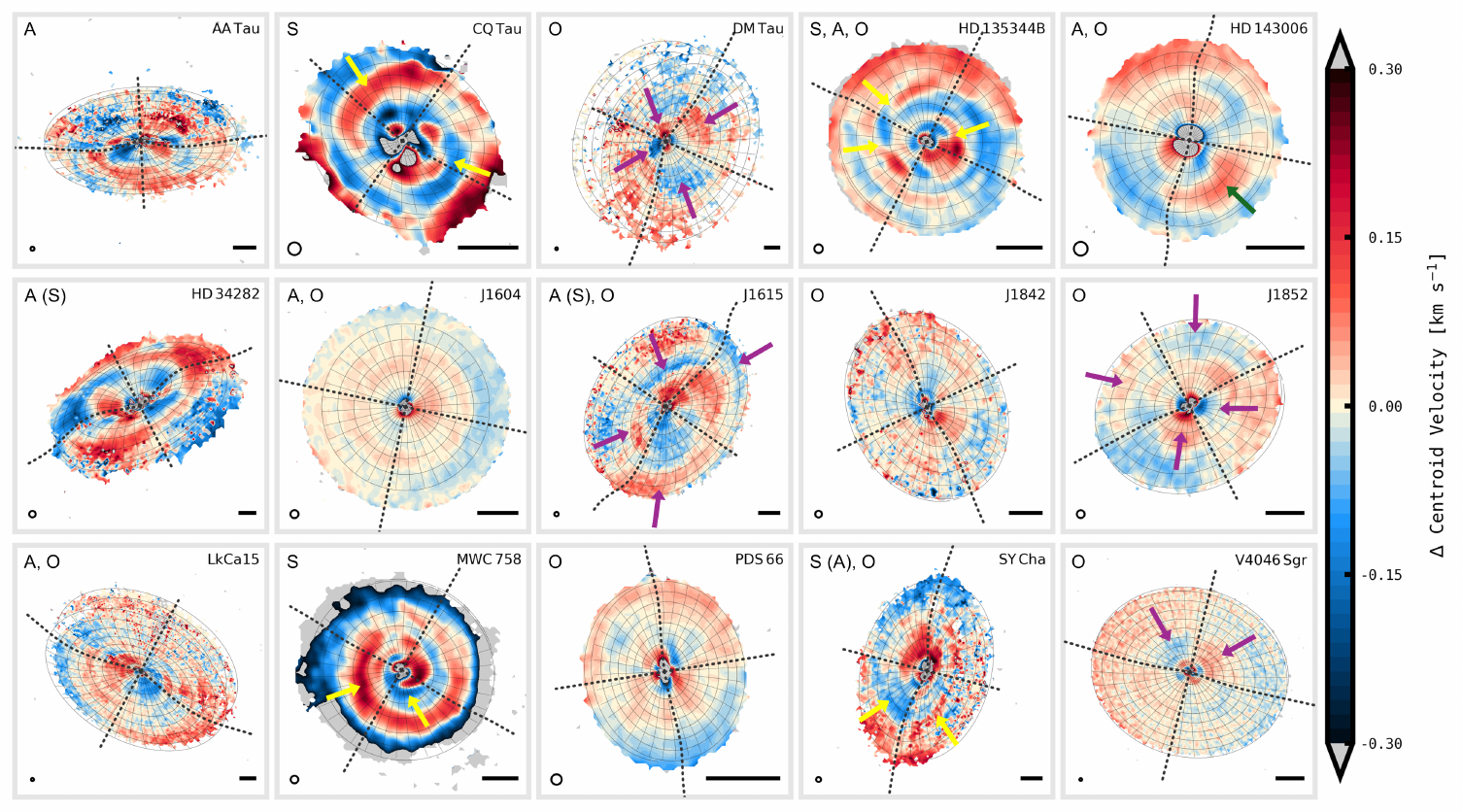}
    \caption{Residual centroid-velocity images of the $^{12}$CO (3-2) emission after subtracting the fitted disk model cube. The definition of lines and beam-size indication are the same as Figure~\ref{fig:velocity_12co}. Pixels with S/N below 4.5 are masked. The yellow, purple, and green arrows indicate the spiral features (Section~\ref{sec:spiral_vres}), the quadrupole pattern (Section~\ref{sec:other_quadrupole}), and the red-shifted arc (Section~\ref{sec:arcs_combined}), respectively. The letter in the upper-left corner of each panel indicates the classification defined in Section 3.2: “S” for spiral, “A” for arcs/rings, and “O” for other types. Labels in parentheses indicate features that appear to be present but are ambiguous.}
    \label{fig:velocity_res}
\end{figure*}

\begin{figure*}
    \centering
    \includegraphics[width=\textwidth]{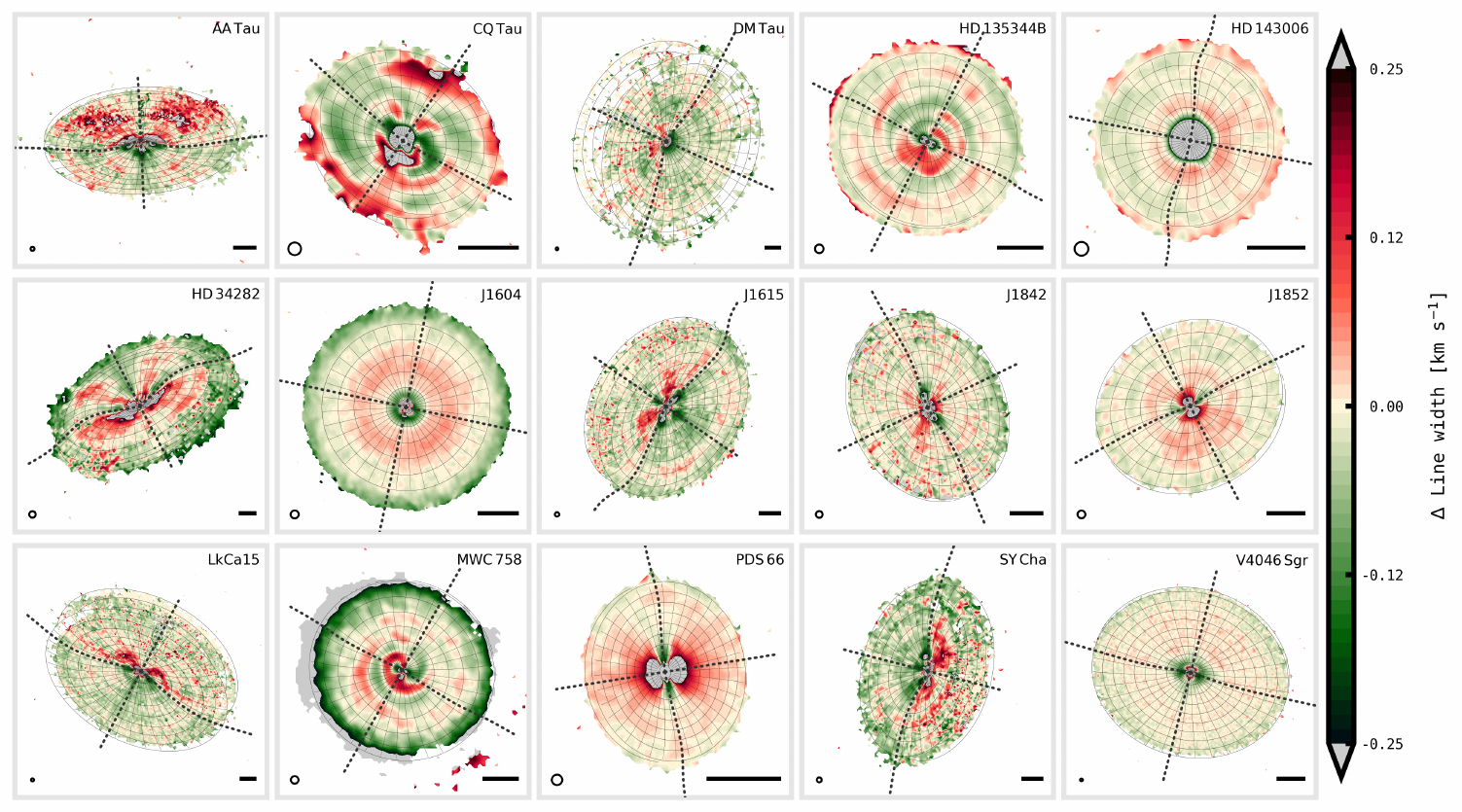}
    \caption{The same as Figure~\ref{fig:velocity_res} but for the line width. Pixels with S/N below 4.5 are masked.}
    \label{fig:linewidth_res}
\end{figure*}

\begin{figure*}
    \centering
    \includegraphics[width=\textwidth]{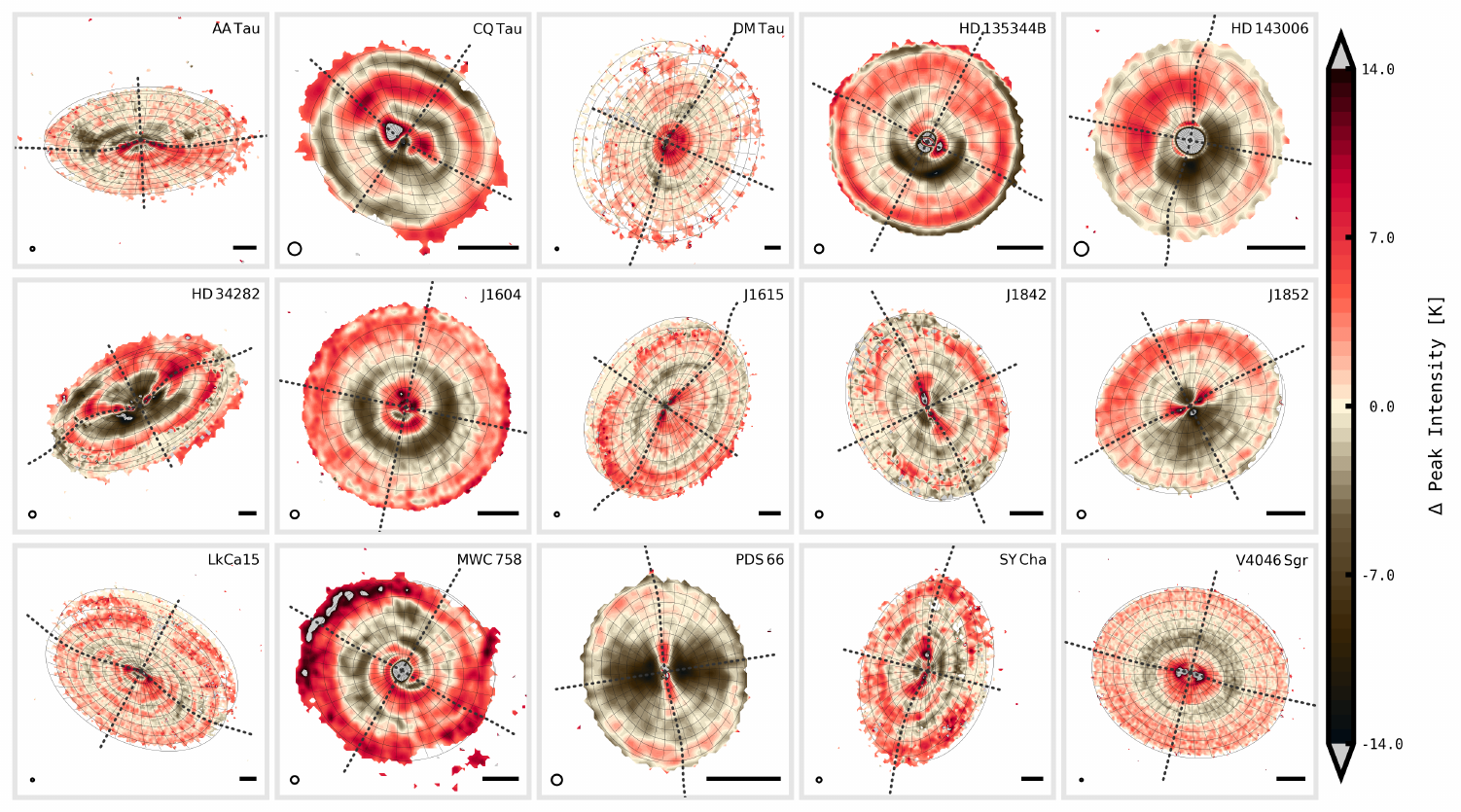}
    \caption{The same as Figure~\ref{fig:velocity_res} but for the peak intensity. Pixels with S/N below 4.5 are masked.}
    \label{fig:peakintensity_res}
\end{figure*}

In the following subsections, we describe the deviations from a smooth Keplerian disk. Residual maps were created by subtracting the model disk obtained from spectral fitting using \textsc{discminer}, and shown in Figures~\ref{fig:velocity_res}, \ref{fig:linewidth_res}, and \ref{fig:peakintensity_res} for the centroid velocity, line width, and peak intensity, respectively. Hereafter, the position angle (P.A.) for the observational data is defined as the angle measured counterclockwise from north to east in the plane of the sky.

First, based on the velocity residuals (Figure~\ref{fig:velocity_res}), we roughly classify the sources into three categories according to the spatial distribution of the residuals: (1) spiral, (2) arc- or ring-like, and (3) other types. 
Our focus is on the global structures, rather than on local non-Keplerian features. We also do not discuss the inner region within approximately $0\farcs3$, because the residuals in such region could be strongly affected by beam dilution and the masking applied over a few beam sizes during the \textsc{discminer} fitting. Gas kinematics in this central region are not straightforward to interpret unless clear substructures are present, such as an inner warp manifested as a velocity twist that is apparent without subtracting the model for J1604 \citep{Mayama2018,Stadler2023,Orihara2023}.
In addition, it needs to be kept in mind that the inspection is based on the morphology projected onto the sky. When the disk inclination is large, identifying substructures becomes more difficult, and contamination from the back side of the disk increases \citep[e.g., AA~Tau,][]{exoALMA_discminer}. 
It should also be noted that the sign of the velocity residuals provides clues as to the underlying velocity structure of the disk. For example, 
assuming that the gas velocity has only an azimuthal component, if the residuals have the same sign as that of the background rotation velocity (i.e., blueshifted or redshifted), it indicates azimuthal motion faster than Keplerian, whereas a residual with the opposite sign implies motion slower than Keplerian (Figure 5 in \citeauthor{Teague2019b} \citeyear{Teague2019b}).

\begin{figure}
    \centering
    \includegraphics[width=0.5\textwidth]{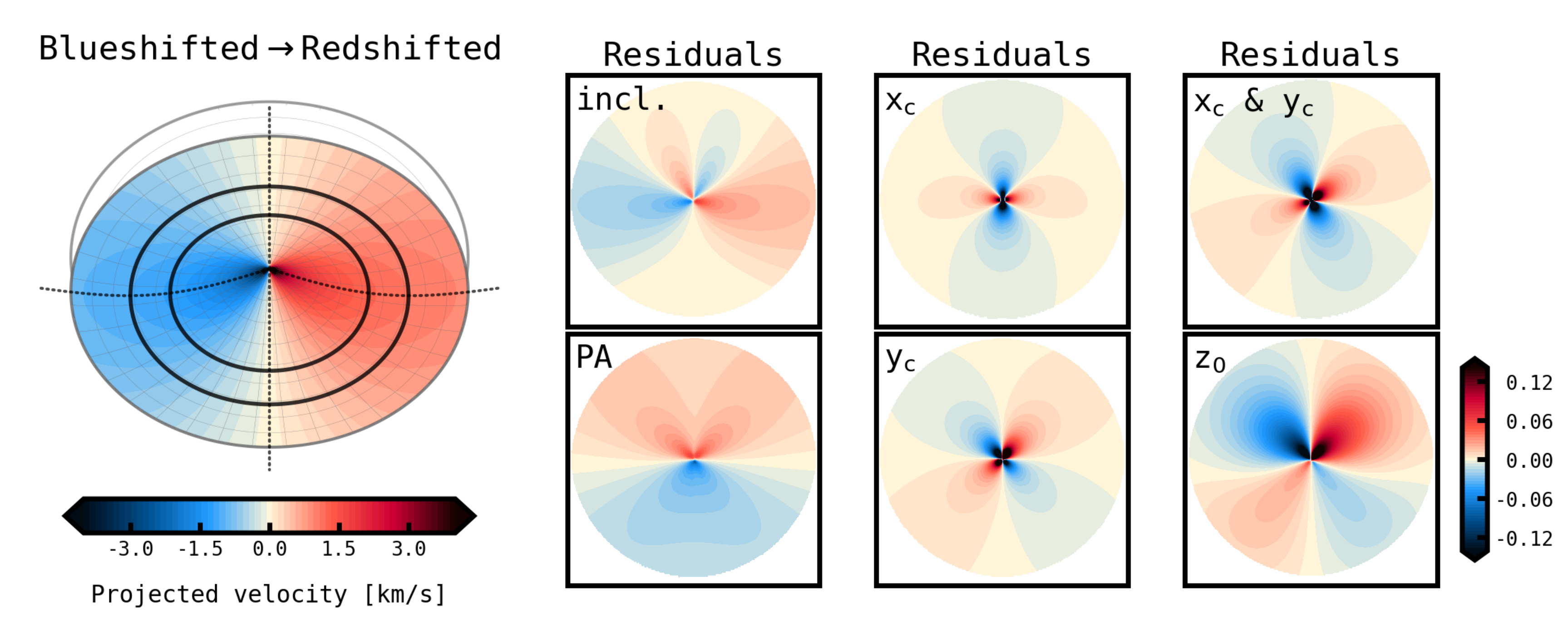}
    \caption{Velocity residuals, shown as a schematic illustration, induced by adding $+1\arcdeg$ to the inclination and P.A., applying a 10\% beam offset to the disk center position in $x_c$, $y_c$, and $x_c+y_c$, and reducing the surface height ($z_0$) by 10\% relative to the ground truth. }
    \label{fig:sketch_error}
\end{figure}

\subsubsection{Spirals}\label{sec:spiral_vres}
Features are regarded as spiral structures when a coherent velocity pattern, possibly changing in sign (blueshifted or redshifted), traces a spiral-like morphology.
CQ~Tau, MWC~758, and SY~Cha fall into this category, showing arms in which the velocity residuals maintain the same sign over large spatial scales (indicated with yellow arrows in Figure~\ref{fig:velocity_res}). HD~135344B shows multiple smaller-scale arms, and with the possible sign flip (Doppler flip), this object can also be classified into this category. As is clearly seen in MWC~758 and HD~135344B, the sign of the residuals along an arm does not change across the disk minor axis, suggesting the presence of a non-azimuthal velocity component and/or a vertical displacement relative to the rotation plane of the bulk gas \citep{exoALMA_vcoherent}.

In CQ Tau, in Figure~\ref{fig:velocity_res}, a redshifted region extends northward from the eastern side at $\sim$100 au, with its position shifting outward in radius toward the north, and appears to continue toward the west. On the western side at $\sim$100 au, a blueshifted velocity component is observed, forming an arm-like pattern inside the red arm. MWC~758 shows a blueshifted region extending from the eastern side at $\sim$30 au, passing through the south and western sides at approximately 120 au. From the eastern side at $\sim$130 au, a redshifted region continues smoothly through the south and west toward the north, where it reaches a radius of roughly 190 au.  In the velocity residuals of SY~Cha, the spiral pattern is less pronounced than in the above two sources. A blueshifted arm is detected on the eastern side at $\sim$290~au, along with a red arm radially just outside it in the southwest.
In HD~135344B, a pattern suggestive of spiral arms is seen across the entire disk. In the eastern region at a radius of about 120 au, the velocity residual alternates from redshifted to blueshifted and back to redshifted as the position angle decreases from 180$\arcdeg$ to 0$\arcdeg$. Inside that region, at $\sim$70 au, a redshifted feature is found on the western side but not in the entire azimuthal ranges.

\subsubsection{Arcs and Rings}
Ring or arc/broken-arm-shaped patterns are visible in the velocity residual maps of AA~Tau, HD~135344B, HD~143006, HD~34282, J1604, J1615, and LkCa~15 (Figure~\ref{fig:velocity_res}). 
Notable examples of systems with relatively well-defined substructures include J1604, which shows a redshifted ring at about 150 au, and LkCa~15, which exhibits multiple ring- or arc-like features \citep{exoALMA_lkca15}. J1615 and possibly HD~34282 could be classified as having spiral-arm-like structures. The presence of substructures is further examined in Section~\ref{sec:combined} together with the residuals in the line width and peak intensity distributions.

\subsubsection{Other Features: Outer Edges}\label{sec:other_vres_outeredge}
When focusing on the outer edges of the disks, a residual velocity gradient along the major axis is seen in several objects (DM~Tau, HD~135344B, HD~143006, J1604, J1842, LkCa 15, PDS 66, SY~Cha, V4046 Sgr), with blue- and redshifted features spatially separated by the minor axis. When compared with the velocity  map in Figure~\ref{fig:velocity_12co}, the sign of the residuals, blue or red, is opposite to that of the bulk gas motion, consistent with a reduced rotational velocity. As described by \citet{exoALMA_pressure}, this significant sub-Keplerian velocity can be attributed to the steep radial decline in pressure toward the outer edge. While this trend is clearly confirmed in the radial profiles of the azimuthal velocity component \citep{exoALMA_pressure, exoALMA_rotation_curve}, it is also evident in these 2D images as a common feature observed in multiple disks, appearing as mirrored blue- and redshifted patterns at the outer edges.

\subsubsection{Other Features: Quadrupole Pattern}\label{sec:other_quadrupole}
Another spatial pattern seen in multiple disks (DM~Tau, J1615, J1852, V4046~Sgr) is a quadrupolar structure, in which blue and redshifted regions alternate azimuthally at a nearly constant radius, as indicated with purple arrows in Figure~\ref{fig:velocity_res}.
In the case of DM~Tau, the direction of the velocity deviation reverses at a radius of approximately 200 au. V4046~Sgr shows the quadrupole residual pattern extending out to about 180 au. In J1615, it is observed along the radial regions at $\sim$450~au and likely at 250 au, while in J1852 similar patterns are seen at $\sim$120 and 210 au. The change in velocity sign in this quadrupole pattern occurs along the minor and major axes of the disk, which indicates a possible connection with the disk geometry. 

It is worth pointing out that non-Keplerian features symmetric with respect to the disk minor axis may originate from a mismatch between the actual emitting surface of the disk and the fitted smooth disk model. 
\citet{Aizawa2025} investigate, both analytically and numerically, how variations in disk inclination, position angle, and surface height influence the fitting of the line-of-sight velocity, also taking into account deprojection errors arising from the disk inclination. They demonstrate that a mismatch in the surface height leads to quadrupolar velocity residuals. Since inclination perturbations affect both the velocity fitting and the deprojection process, it does not yield a perfectly symmetric quadrupole, although a similar pattern emerges. In contrast, a perturbation in P.A. produces a bipolar shift in the residuals, with blue and redshifted regions separated along the disk major axis. 
Figure~\ref{fig:sketch_error} presents an example of the velocity residuals induced by slight perturbations in these parameters, offering a schematic view of how these effects can appear in the data.

Compared to the radial profile of the emission-surface height averaged over the azimuthal direction \citep{exoALMA_surface}, the radial ranges exhibiting quadrupole patterns in the above four disks roughly coincide with the radii where the surface height changes discontinuously.
The quadrupole residuals nonetheless suggest that the emitting surface is not radially smooth.

\subsection{Combined Characteristics of Line Width and Peak Intensity}\label{sec:combined}
\begin{figure*}
    \includegraphics[width=0.49\textwidth]{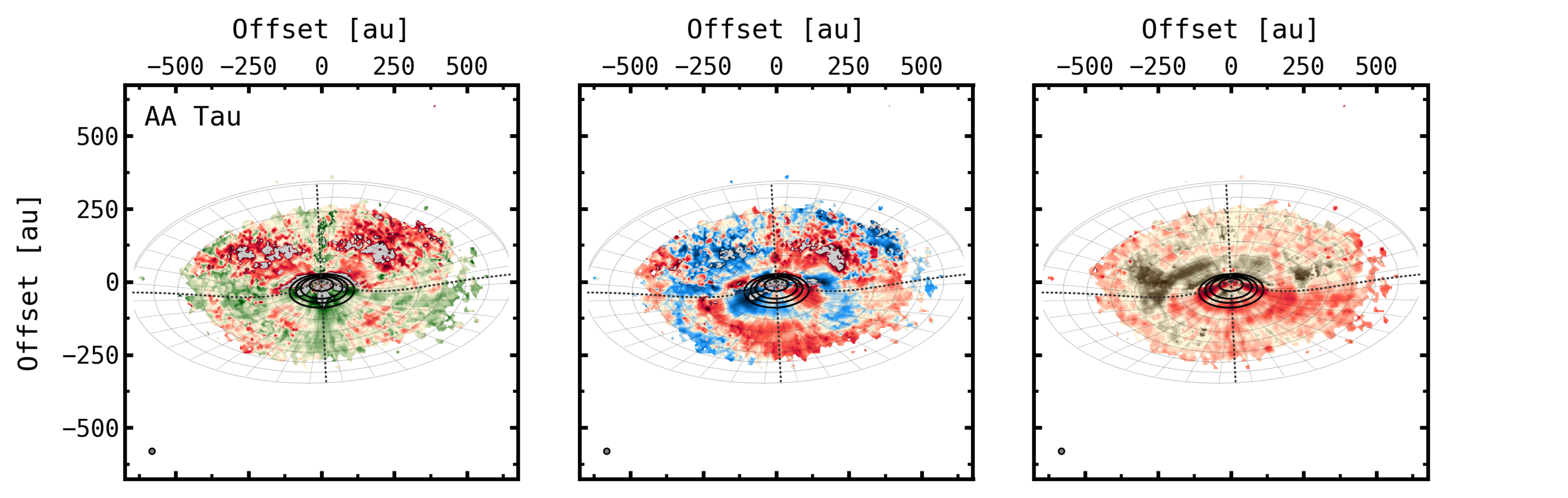}
    \includegraphics[width=0.49\textwidth]{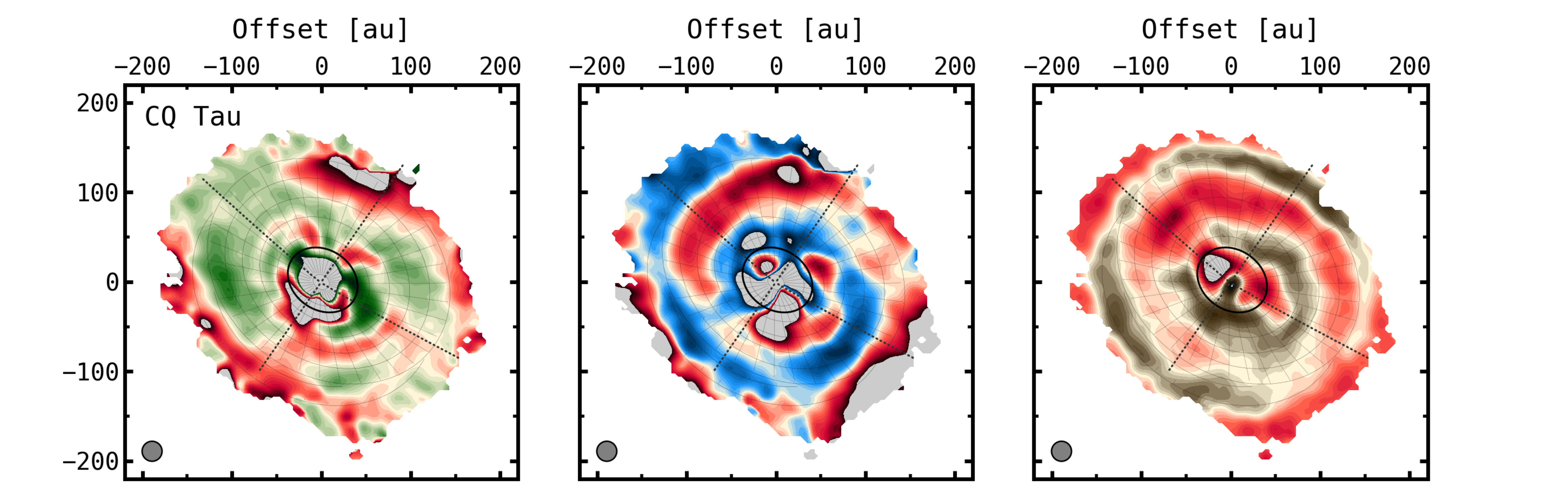}\\
    \includegraphics[width=0.49\textwidth]{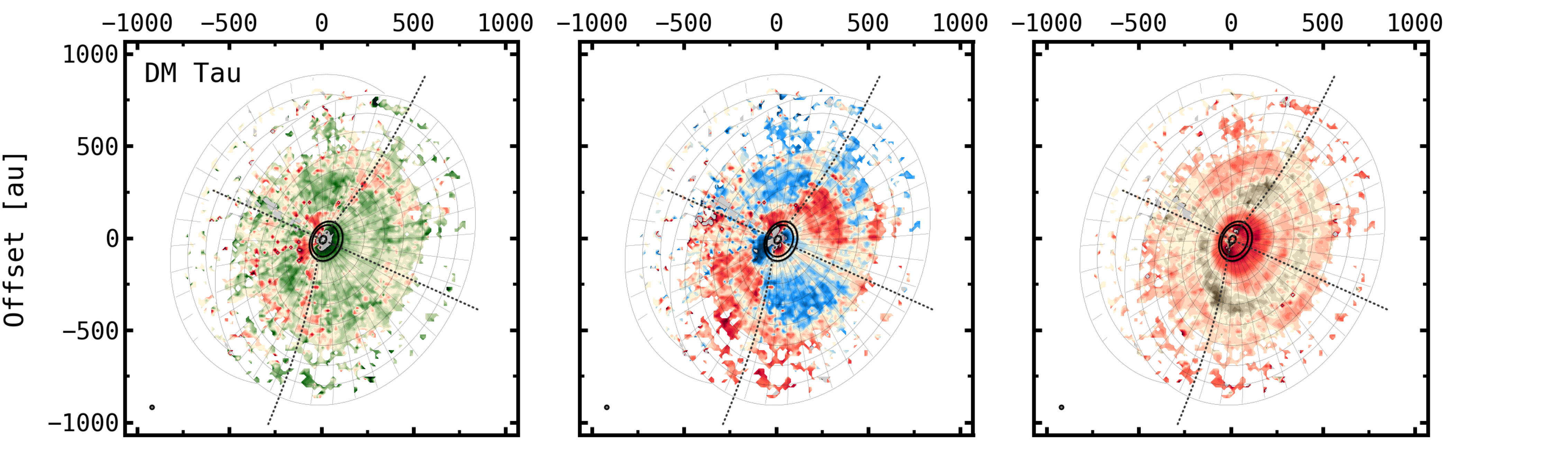}
    \includegraphics[width=0.49\textwidth]{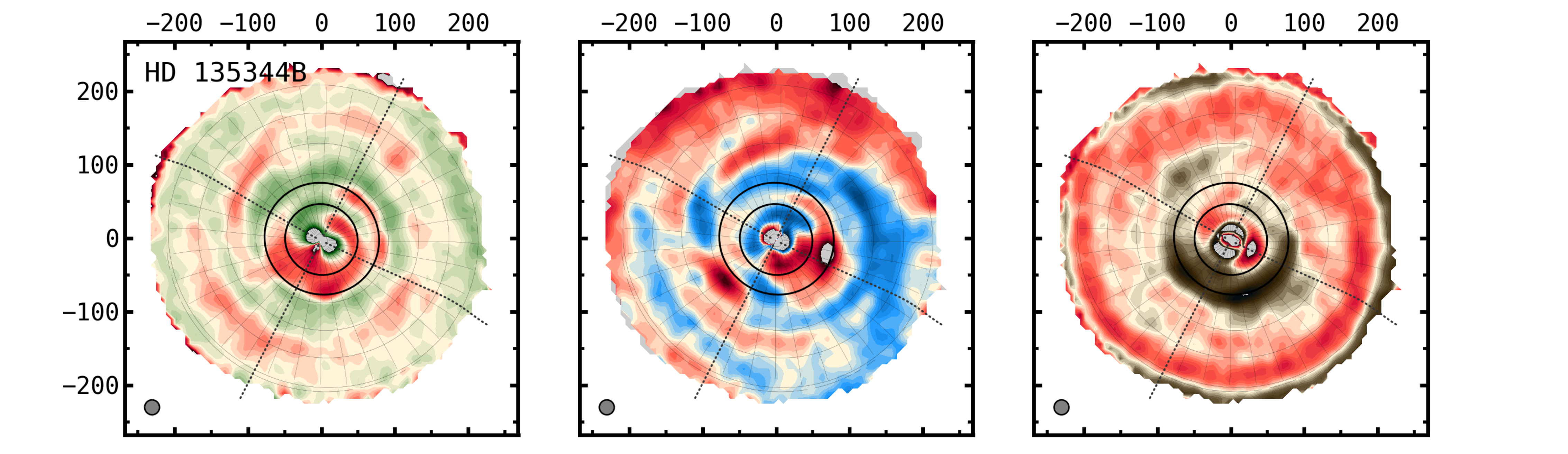}\\
    \includegraphics[width=0.49\textwidth]{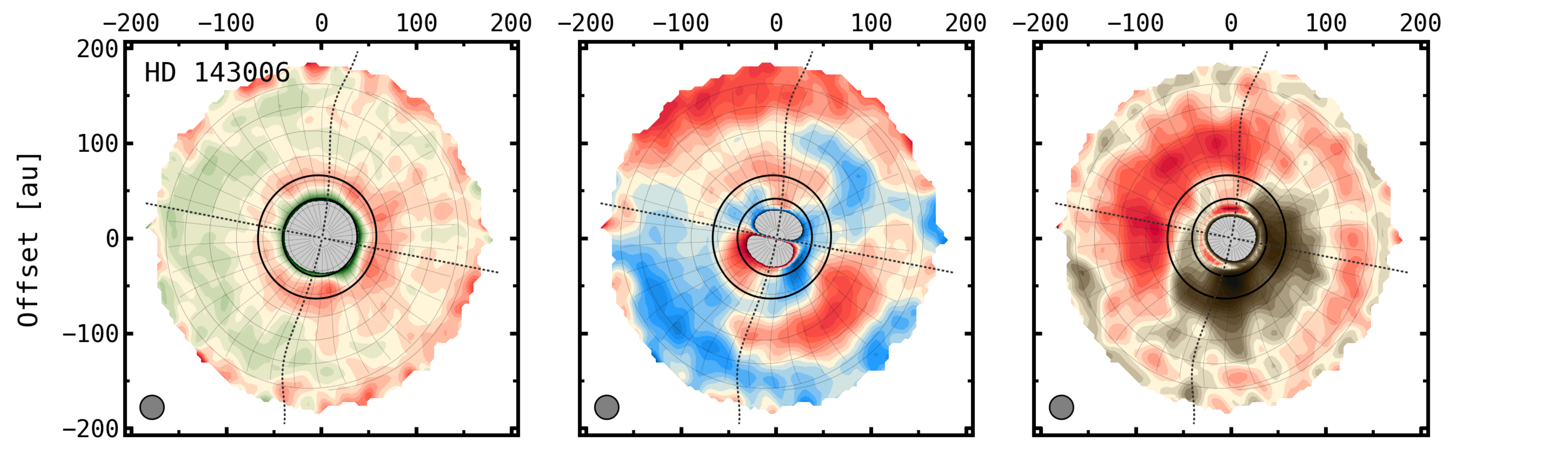}
    \includegraphics[width=0.49\textwidth]{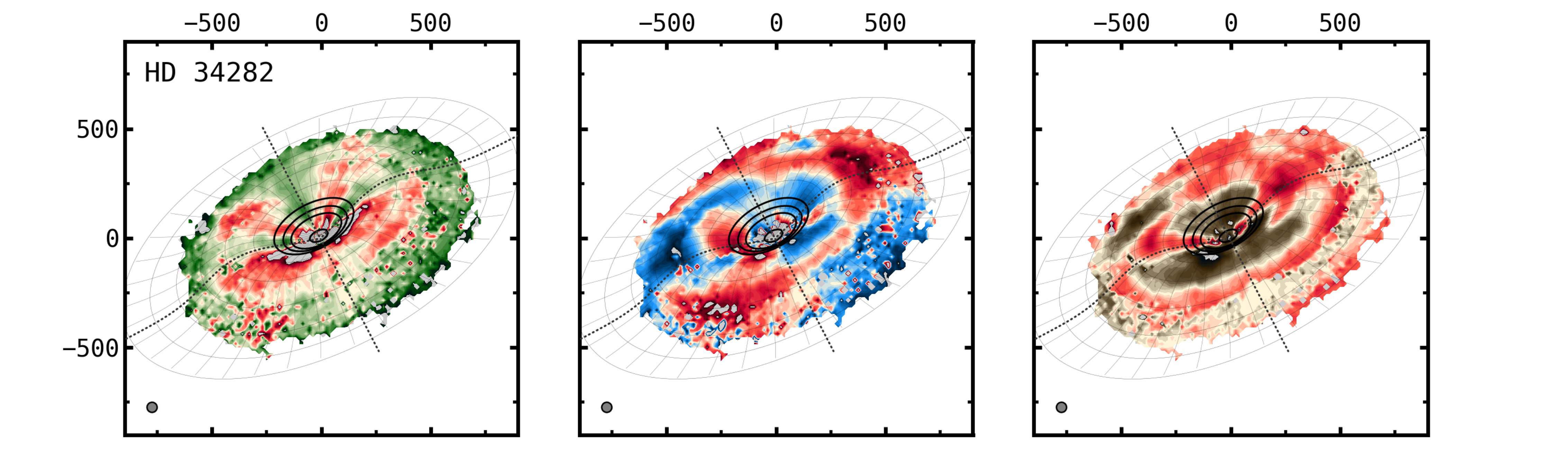}\\
    \includegraphics[width=0.49\textwidth]{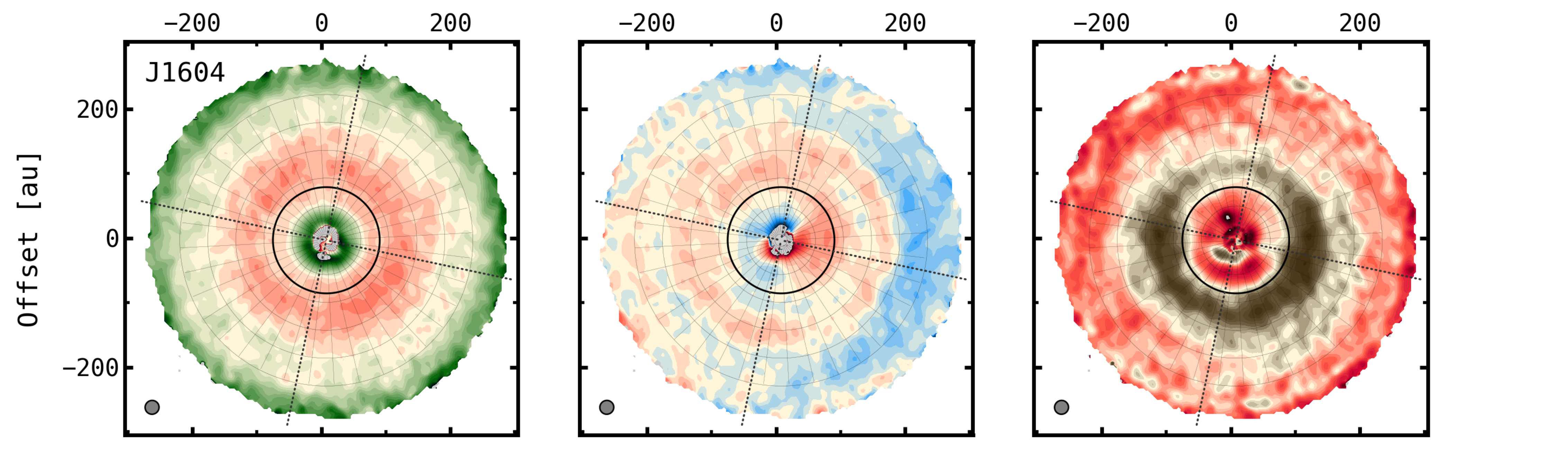}
    \includegraphics[width=0.49\textwidth]{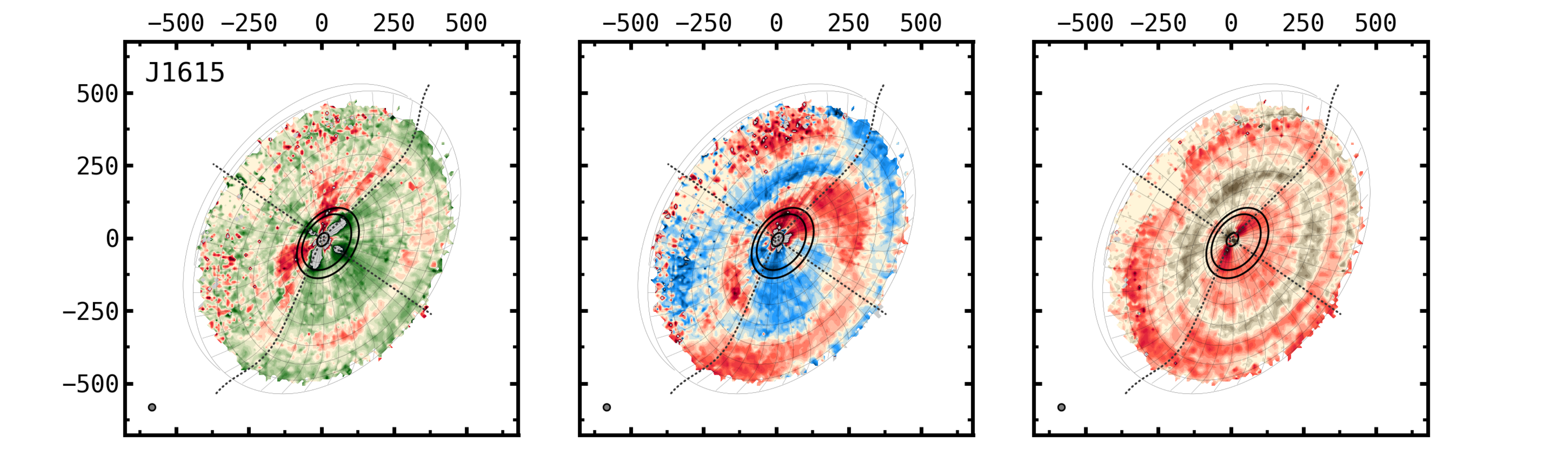}\\
    \includegraphics[width=0.49\textwidth]{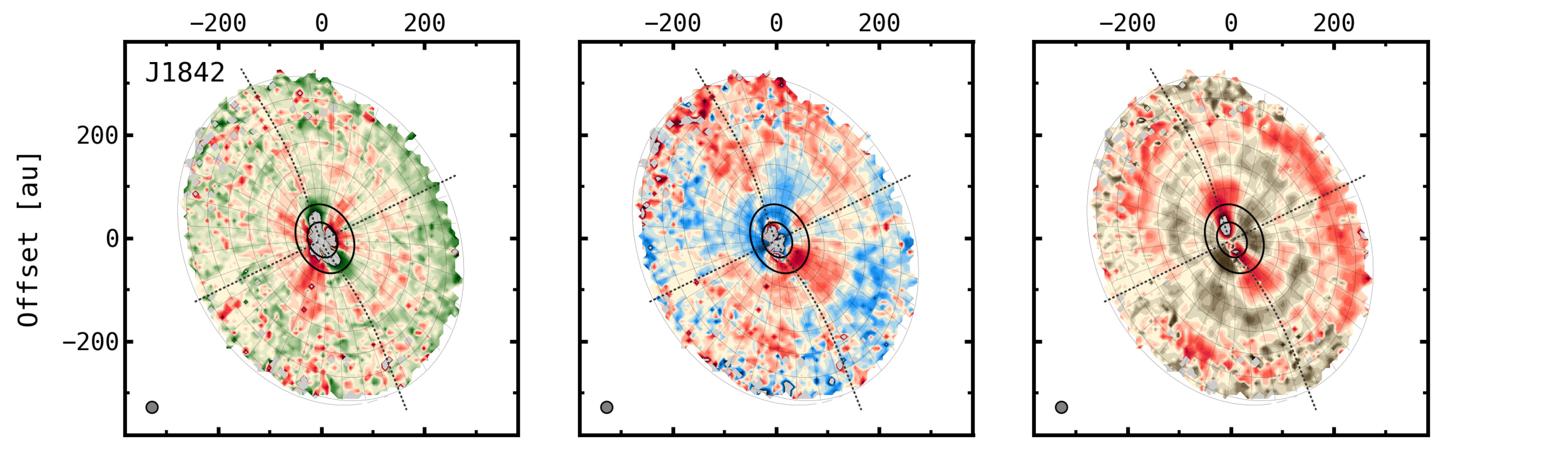}
    \includegraphics[width=0.49\textwidth]{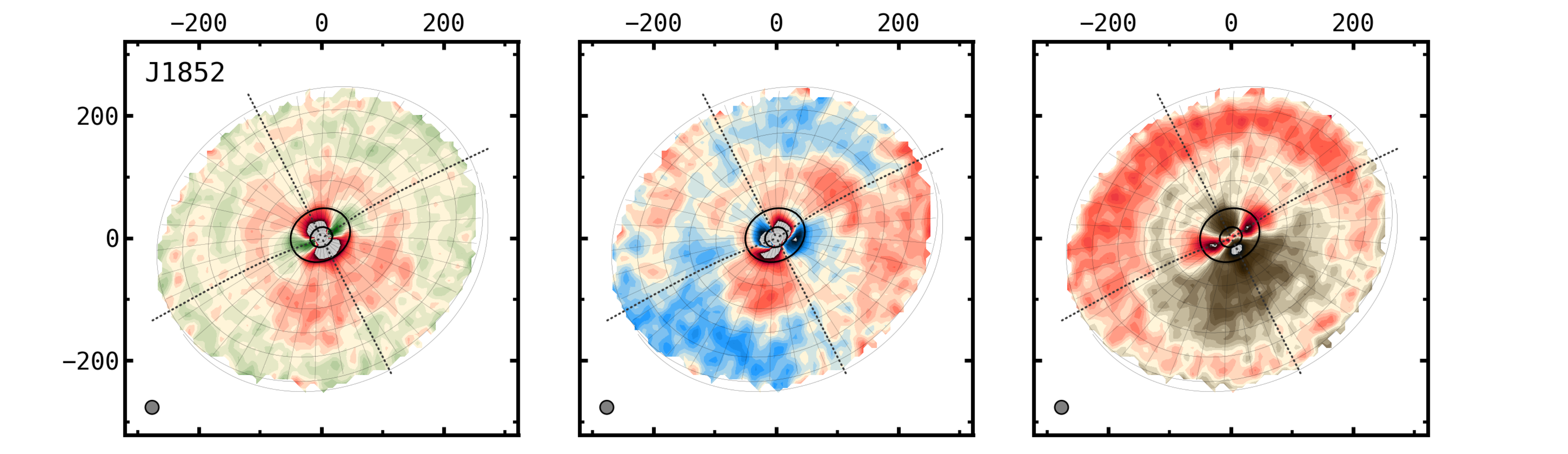}\\
    \includegraphics[width=0.49\textwidth]{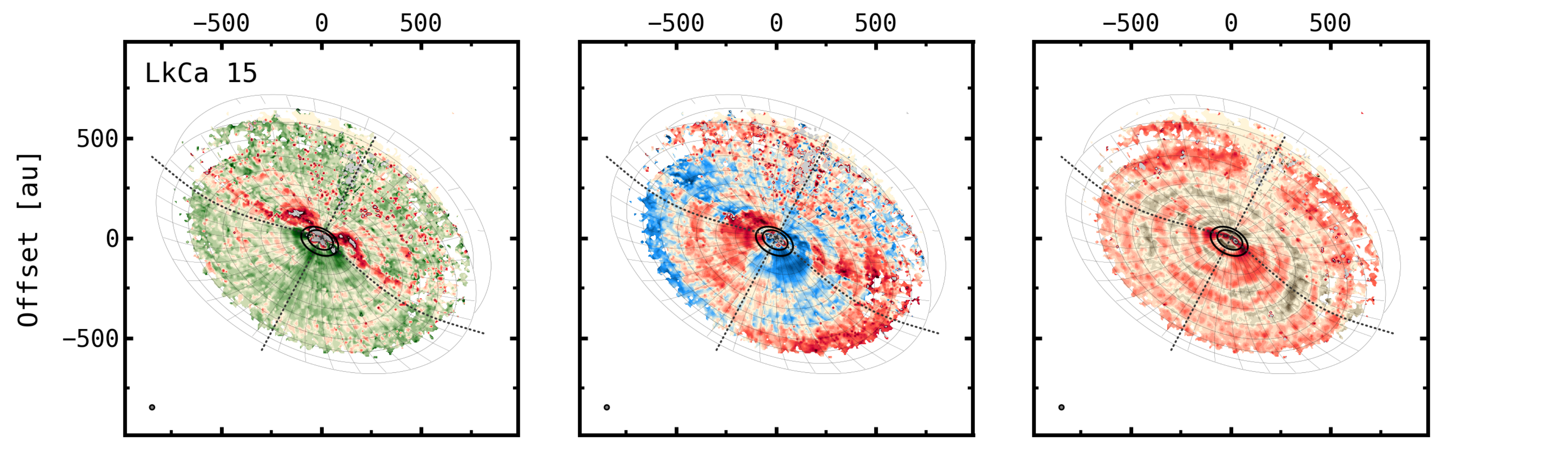}
    \includegraphics[width=0.49\textwidth]{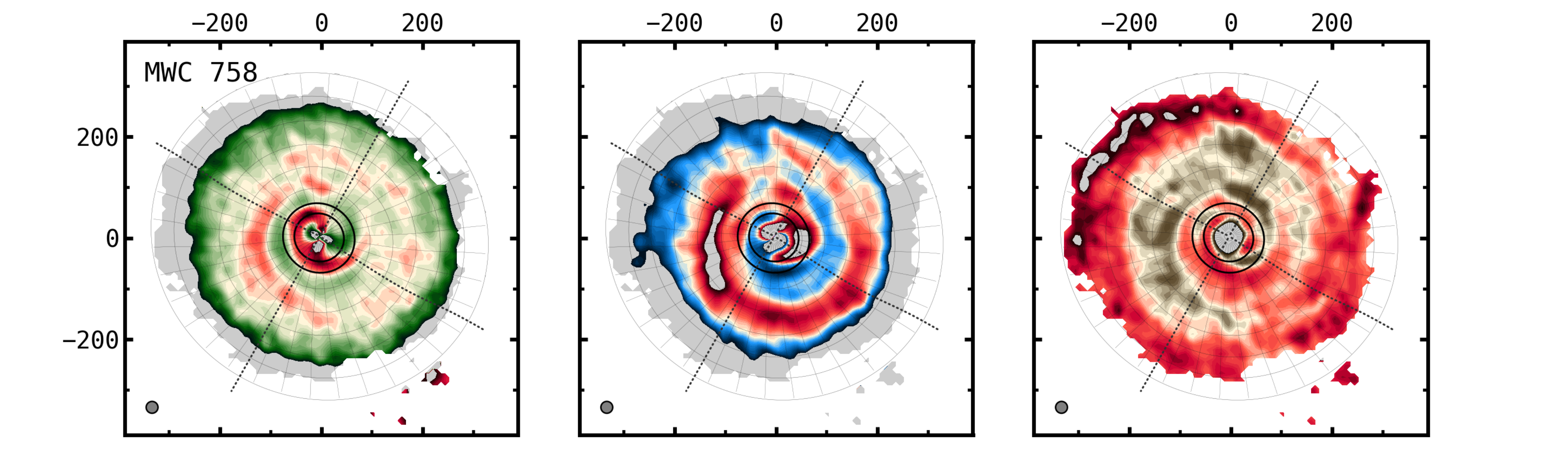}\\
    \includegraphics[width=0.49\textwidth]{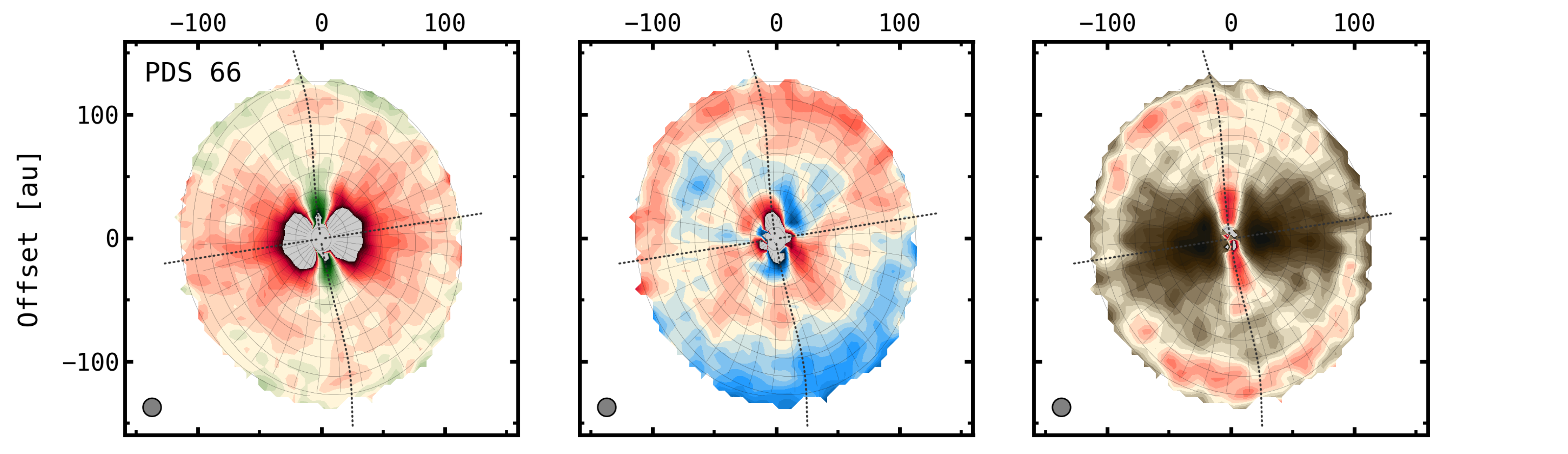}
    \includegraphics[width=0.49\textwidth]{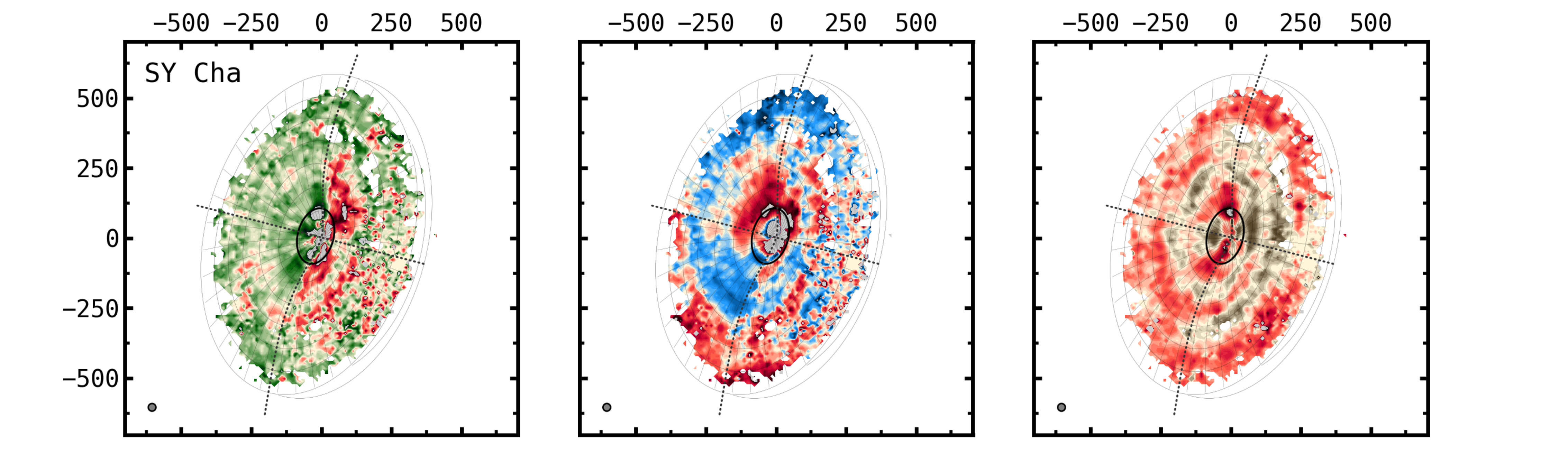}\\
    \includegraphics[width=0.49\textwidth]{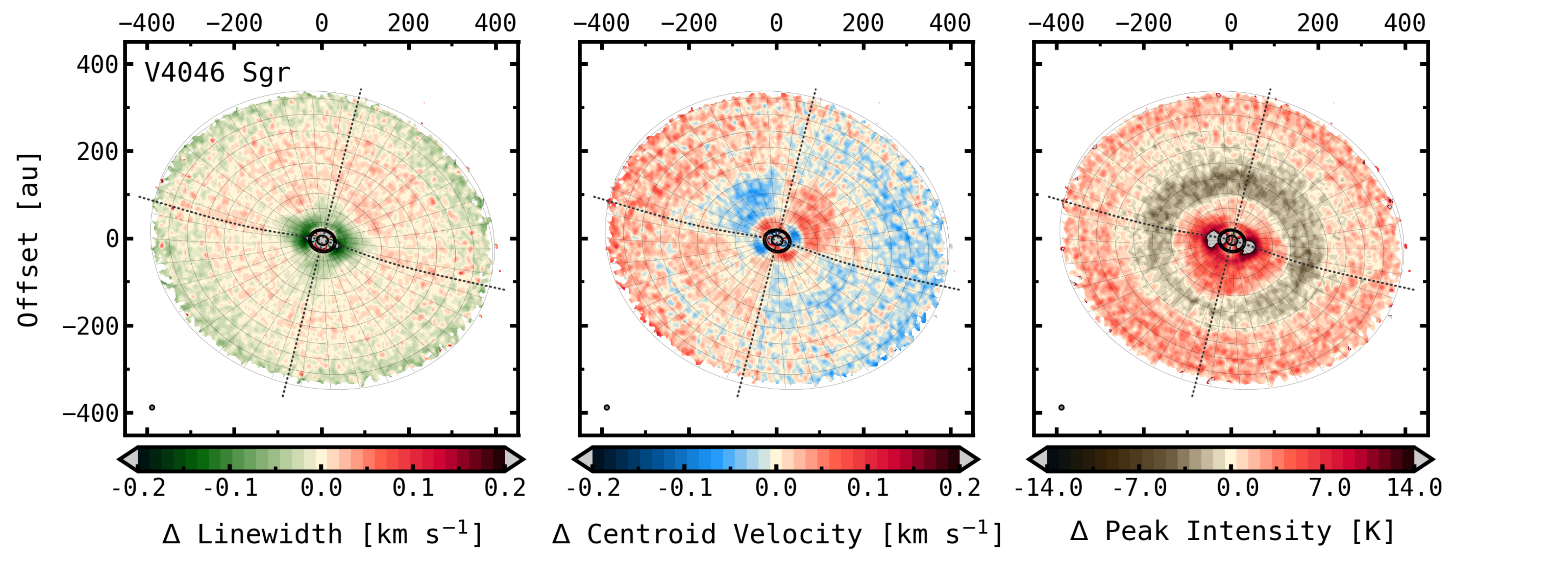}
    \caption{Residuals in line width (left), velocity (middle), and peak intensity (right) after subtracting the fitted Keplerian disk model for each source. The same set of color bars is used for all sources and is displayed only for the lower-left source (V4046 Sgr). A 5-sigma clipping was applied to generate the moment maps. The midplane locations of the dust continuum rings, projected onto the CO emission surface, are shown as black ellipses \citep{exoALMA_dust}.  }\label{fig:residuals_all}
\end{figure*}

In this subsection, we extend the discussion of the velocity residual morphology by examining the residuals in the line width and peak intensity. Figure~\ref{fig:residuals_all} present the residual maps arranged side-by-side to enable comparison of these 3 residuals for each object.
Table~\ref{tab:summary} summarizes the features found in these images.

\subsubsection{Spirals}
MWC~758, CQ Tau, and HD~135344B, classified as having spiral structures in the previous subsection (Figure~\ref{fig:velocity_res}), show the spiral-shaped or arm-like morphology also in the line width and peak intensity. 
For MWC 758, along the 30--120 au arm seen in the blueshifted velocity residuals, the peak intensity is generally positively offset. The line width is broader in the same region on the eastern side of the disk and along the inner edge of the blueshifted arm from the south to the west. Regarding the redshifted arm at 130--190 au, the spatial correspondence in the peak intensity residuals is not as clear; the eastern side shows lower intensity, while the western side exhibits somewhat higher intensity than the fitted smooth model. On the other hand, the line width increases along the redshifted arm.

In CQ~Tau, the prominent arm extending eastward from $\sim$100 au appears as a positive feature in the peak-intensity residuals. The line width is enhanced along the outer edge of the northern velocity arm. In addition, an additional arm-like region of broad line width is present on the southwestern and southern sides at $\sim$100 au. This feature coincides with the radial boundary where the velocity residuals switch from red to blue and where the peak-intensity residuals change the sign. It is likely that the line width becomes broader in regions where the radial velocity gradient is large. 

In the case of HD~135344B, the velocity flip from blue to red observed at 110 au and P.A. of 40$\arcdeg$ corresponds to the characteristic deficit in peak intensity seen toward the northeast. Another velocity flip, at 90 au and 100$\arcdeg$, is located at the boundary where the residuals in peak intensity change sign from negative to positive.
The line width becomes broader along the outer edge of the arm where these velocity flips occur, corresponding to the outer edge of the gap seen in the peak intensity. 
The line-width residual map shows an arm-like region of broad line width extending from the west ($\sim$140 au) to the south ($\sim$180 au), which coincides with the radial boundary between blue and red velocity residuals.
The regions with enhanced line width correspond to radial locations where the velocity residuals change sign as is also seen in arm-like structures in other disks. 

These three objects exhibit spiral-like similarities in velocity, line width, and peak intensity. The line width, which is insensitive to the sign of the velocity offset, can be a useful tool for tracing characteristic velocity gradients. A comparable spatial correlation between line width and velocity offset is also observed, although the deviation in line width is marginal ($<$0.1 km s$^{-1}$), on the southwestern side at 290 au in SY~Cha, in the western side at 500~au in HD~34282, and in the western half at 350 au in J1615. 

\subsubsection{Arcs and Rings}\label{sec:arcs_combined}
All objects showing arc/broken-arm or ring-like structures in velocity (AA~Tau, LkCa~15, HD~34282, J1604, J1615), except HD~143006, exhibit similar substructures in the peak intensity residuals, where the features are even easier to recognize. In addition, J1842 shows positive ring-like residuals at $\sim$270 au.
Although the contrast is weak, in the peak intensity residuals, DM~Tau also has multiple rings including the arc on the western side (P.A.$\sim$$210\arcdeg$--$290\arcdeg$) at $\sim$250~au and a tenuous ring/arc at about 450--550 au \citep{exoALMA_dmtau}.  

Conversely, the correspondence with the line width residuals is not so obvious. An arc-like structure can be identified in HD~34282, SY~Cha, and J1615, as noted earlier. There might be substructures that appear as arcs or broken rings also for AA~Tau and LkCa~15, although these are not significant in the sense that their amplitudes approach the velocity resolution limit of the data.  

HD~143006 exhibits a feature that is not commonly seen in the other sources. The peak intensity residuals show the global mismatch where the far-side of the disk is brighter than the near-side in observations than in the model, both in $^{12}$CO and $^{13}$CO. Indeed, \citet{exoALMA_dust} find that the eastern side is brighter in dust continuum, consistent with the view that the eastern side has the higher temperature. This can be explained by the asymmetric illumination of the outer region as first detected in scattered light, indicative of shadows cast by a misaligned inner disk structure \citep{Benisty2018}. 
In the velocity residuals image, it shows a redshifted arc at a radius of $\sim$100 au over a P.A. range of 160$\arcdeg$--240$\arcdeg$ (indicated with a green arrow in Figure~\ref{fig:velocity_res}). When compared with previous studies \citep{Perez2018,exoALMA_dust,exoALMA_asymmetries}, the redshifted arc in $^{12}$CO is located just outside the 64-au dust ring in the radial direction. 
In $^{13}$CO, this region lies near the outer boundary of the area with sufficient S/N for discussion, and it appears redshifted and is therefore not totally inconsistent with the $^{12}$CO distribution. However, no significant residuals are found in the $^{12}$CO line width at the same or similar location, favoring an interpretation other than gas kinematics.
Given that this redshifted arc occupies one quadrant defined by the minor and major axes of the disk, it may be part of a quadrupole pattern that can arise from a mismatch in the height of the gas-emitting surface (Section~\ref{sec:other_quadrupole}). For this particular object, however, azimuthal temperature variations by the shadows may instead provide a more plausible explanation, as similar quadrupole patterns are seen in hydrodynamical simulations of shadow-driven temperature substructures \citep[e.g.,][]{Zhang2024}. 

In the peak intensity residuals for J1604, there is a deficit just outside the dust continuum ring up to $\sim$150 au in $^{12}$CO, and also in $^{13}$CO. At this boundary of negative and positive residuals in the peak intensity at 150 au, the line width is larger than the model and the centroid velocity tends to be redshifted in $^{12}$CO and $^{13}$CO. The sign of the velocity shift does not flip at the disk minor axis, suggesting the presence of a vertical (non-azimuthal) velocity component along this ring \citep{Stadler2023, exoALMA_vcoherent}. 

\subsubsection{Other Features}\label{sec:other_axis-dependent}
PDS~66, J1842, and J1852 show residuals where the peak intensity is higher and the line width is narrower along the major axis, while the opposite trend appears along the minor axis. A similar, though weaker, pattern is seen in $^{13}$CO compared to $^{12}$CO. This pattern can be explained by beam dilution in an inclined disk, as the major axis is least affected by this effect. Another possible explanation, as mentioned in \citet{exoALMA_discminer}, is optical depth variation when the line is not highly optically thick. The optical depth depends on column density divided by line width (Figure~\ref{fig:linewidth_res}); in an inclined rotating disk, it tends to be larger along the major axis and smaller along the minor axis. Consequently, the major axis may appear brighter either because we observe higher, warmer layers as the optical depth reaches unity more easily in this region compared to the minor axis. This effect is expected to be insignificant when the optical depth is much greater than one. The three disks lie at the lower end of the disk-to-star gas mass ratio estimated by \citet{exoALMA_rotation_curve}, consistent with this interpretation. 
Therefore, for these 3 objects, residuals arising from the inclination effect may dominate, leaving open the possibility that signatures of substructures are buried and thus remain undetected. Nevertheless, J1842 exhibits a ring morphology at the outer edge of $\sim$270~au in peak intensity (Figure~\ref{fig:peakintensity_res}), while other two show no such noticeable substructures. In J1852, the far side of the outer region appears brighter than the near side in the peak-intensity residuals, similar to that in HD~143006, but no shadow has been reported in scattered light \citep{Villenave2019}.

No substructure is detected in the linewidth residuals for V4046 Sgr. 
Meanwhile, the peak intensity exhibits a decrease at a radius of $\sim$150 au, appearing more prominently on the near-side with a wider radial range than on the far-side to us, both in $^{12}$CO and $^{13}$CO lines. 
This intensity deficit has been captured in \citet{exoALMA_surface}, where the height of the emission surface has a double peak, with the lowest surface height occurring at 173 au between the two peaks. \citet{exoALMA_surface} discuss the possible cause of this feature through the inspection of channel maps and comparison with a disk model, implying that it is likely caused by projection effects, possibly associated with low optical depth or low temperature.

In PDS~66, while substructures are difficult to extract in the line width and peak intensity due to the reason described above, no significant features are detected in the velocity field, except for the sub-Keplerian rotation at the outer edge. This quiescent kinematic structure is consistent with the smooth appearance of the dust continuum emission at 0.9~mm \citep{exoALMA_dust} and in the $^{12}$CO channel maps \citep{exoALMA_kink}, but does not necessarily preclude substructure formation in larger grains, as seen in the 3 mm continuum \citep{Ribas2025}.

\begin{deluxetable*}{lcccccl}
\tabletypesize{\scriptsize}
\tablecaption{Features in the residuals images in $^{12}$CO\label{tab:summary}}
\tablehead{
\colhead{Source} & \colhead{Spirals} &  \colhead{Arcs/rings$^{1}$} & \colhead{\shortstack{Outer-edge\\sub-Keplerian}} & \colhead{\shortstack{Quadrupole\\in  velocity}$^{2}$} &  \colhead{Dust continuum$^{3}$}  & \colhead{Comment}
}
\startdata
AA Tau     & -             & \checkmark         & -          &             & rings, shadows            & \\
CQ Tau     & \checkmark    & -                  & -          & -           & spirals                   & \\
DM Tau     & -             & \checkmark (in $T$) & \checkmark  & \checkmark  & rings, inner disk offset & \\
HD 135344B & \checkmark    & \checkmark         & \checkmark &  -          & ring, arc                 & \\
HD 143006  & -             & \checkmark         & \checkmark & -           & rings, arc, shadows        & a redshifted arc at $\sim$100 au and P.A. of 160$\arcdeg$--240$\arcdeg$\\
HD 34282   & (\checkmark)  & \checkmark         & -          & -           & rings, arc            & \\
J1604      & -             & \checkmark         & \checkmark & -           & ring, shadows             & ring at $\sim$150 au always redshifted\\
J1615      & (\checkmark)  & \checkmark         & -          & \checkmark  & rings, inner disk offset  & \\
J1842      & -             & \checkmark (in $T$) & \checkmark & -          & rings, shadows              & axis-dependent brightness$^{4}$ \\
J1852      & -             & -                  & -          & \checkmark & rings                      & axis-dependent brightness$^{4}$   \\
LkCa 15    & -             &  \checkmark        & \checkmark & -           & rings                     & \\
MWC 758    & \checkmark    & -                  & -          & -           & rings/arcs                & \\
PDS 66     & -             & -                  & \checkmark &  -          & smooth                    & axis-dependent brightness$^{4}$ \\
SY Cha     & \checkmark    & (\checkmark)       & \checkmark & -           & ring                  &\\
V4046 Sgr  & -             & \checkmark (in $T$) & \checkmark  & \checkmark & rings, inner disk offset   & deficit in peak intensity at $\sim$200 au\\ 
\enddata
\tablecomments{
Checkmarks in parentheses indicate features that appear to be present but are difficult to identify clearly.
1. "(in $T$)" means that the feature is seen only in the peak intensity.  
2. This velocity residual pattern may not be of dynamical origin, as it can result from a mismatch between the observed and modeled heights of the CO emitting surface. Even if this is the case, radial fluctuations of the emitting surface relative to the underlying disk would indicate physical deviations from a smooth Keplerian disk (Section~\ref{sec:other_quadrupole}). 
3. \cite{exoALMA_dust}. 
4. The brightness pattern along the major and minor axes of the disk may be affected by optical depth effects along the line of sight. Because this pattern is strong, it is possible that finer substructures in brightness are not fully detected. See Section~\ref{sec:other_axis-dependent}.
}
\end{deluxetable*}

\subsection{Overview of the Results}
The residual maps for centroid velocity, line width, and peak intensity reveal deviations from smooth Keplerian disks in all sources (Table~\ref{tab:summary}). Some disks exhibit spiral structures in which the spatial distributions of residuals are correlated among centroid velocity, line width, and peak intensity, while others display arc-like or ring-like structures in which substructures are more readily visible in peak intensity. In addition, quadrupole patterns are present in the velocity residuals in several sources. These may be interpreted not as dynamical structures, but as deviations in the height of the emitting surface or as azimuthal temperature variations. On the other hand, in peak intensity, residual patterns aligned with the disk minor and major axes are also observed. These can be caused by inclination and optical depth effects, and therefore may not necessarily reflect the actual density or temperature structures. 
No clear correlation is found between the above characteristics and stellar properties such as mass and age, except for a tendency that Herbig Ae/Fe disks exhibit a higher degree of non-axisymmetry than T Tauri disks. Regarding disk properties, vertically flatter structures and lower disk masses are suggested to be associated with weaker non-axisymmetric features, as described below.

The sources that show no spatially noticeable substructures in the line-width residual maps, excluding the increase in line width along the disk minor axis due to line-of-sight optical depth variations, are DM Tau, HD 143006, J1842, J1852, PDS~66, and V4046~Sgr. In Table~\ref{tab:summary}, these tend to be the systems that are not flagged as hosting spirals or arcs/rings. Except for HD 143006, which may exhibit shadow-induced temperature variations, J1852, PDS~66, and V4046~Sgr show no non-axisymmetric features or prominent arc/ring structures in their peak intensity images. Nevertheless, in the velocity residuals, J1852 and V4046~Sgr show quadrupole patterns, indicating that deviations from smooth disks are still present in these systems. It is worth noting that J1852, PDS~66, and V4046~Sgr have vertically flatter disks and lower disk-to-stellar mass ratios, below 5\%, according to disk masses estimated by \cite{exoALMA_rotation_curve} for 10 exoALMA sources from the rotation curves with gas self-gravity taken into account (see also \citeauthor{exoALMA_gassmass2} \citeyear{exoALMA_gassmass2} for mass estimates). Furthermore, these three sources show the smallest non-axisymmetry indices in dust continuum \citep{exoALMA_dust}.

In contrast, CQ~Tau, MWC~758, and HD~135344B, which host spiral structures, show significant non-axisymmetric gas motions. In addition, HD~34282 suggests arc/arm-like features that correspond to the arcs/spiral arms imaged in near-infrared scattered light \citep[][]{deBoer2021}. 
If HD~34282 indeed has a spiral structure, although it is not easy to discern because of its relatively large inclination, spirals would then be present in all Herbig Ae/Fe stars in the exoALMA sample. In addition, in these disks, large-scale azimuthal asymmetries have also been detected in the dust continuum emission \citep[e.g.,][]{exoALMA_dust,Cazzoletti2018,Casassus2019, Kuo2024}. 
SY~Cha lies in the T~Tauri regime, and the degree of non-axisymmetry is indeed smaller than that in the disks of these Herbig Ae/Fe stars, as is also seen in the dust continuum emission \citep{exoALMA_dust}.
Outside the exoALMA program, globally coherent wiggles in the gas kinematics have been detected in AB Aur, strongly suggesting gravitational instability \citep{Speedie2024}. Although the mechanism responsible for driving spiral structures need to be investigated independently for each source, AB Aur is also classified as a Herbig Ae star.
While stellar masses (i.e., gravitational effects) are generally larger and disk temperatures are higher for Herbig Ae/Fe stars than T Tauri stars, these differences do not directly imply a higher degree of non-axisymmetry. The overall disk structure may be linked to the initial disk mass or to the duration and amount of material accreted from the surrounding environment.

\citet{Winter2025} demonstrate that disk warps can cause large-scale velocity residuals, leading to a spiral-like residual pattern, by analyzing the slight mismatch in the inclination and P.A. from those of a smooth disk. \citet{Aizawa2025} also show through analytical modeling that the effects of perturbations in inclination, P.A., and the height of the emitting surface are decomposed into azimuthal modes with $m$$=$1–-3. Since deviations in inclination and the surface height exhibit (anti-)symmetry with respect to the disk minor axis, their contributions can be suppressed by subtracting the mirror image across the minor axis.
\cite{exoALMA_discminer2} and \cite{exoALMA_vcoherent} applied a folding analysis to the exoALMA data, in which the residuals from a smooth Keplerian disk are compared between the two halves divided by the disk minor axis and axisymmetric residuals are filtered out. This methodology enhances the robustness of the scientific interpretation related to the extraction of non-axisymmetric features induced by embedded planets or other mechanisms. Detailed analyses and discussions of individual sources are presented in those papers. Further dedicated studies of localized non-Keplerian substructures will be presented as a future work (S. Facchini et al. in prep.).

\section{Summary}
We uniformly analyzed two-dimensional maps of centroid velocity, line width, and peak intensity in CO for 15 disks, with an angular resolution of $0\farcs15$ and a velocity resolution of 100 m s$^{-1}$. As a result, all targets show deviations from smooth Keplerian disks in terms of gas kinematics and/or intensity.

\begin{itemize}
\item The disks around five sources (CQ Tau, MWC 758, HD 135344B, SY Cha, and likely HD 34282) show spiral-like arms in the residuals of velocity, line width, and peak intensity, which tend to be preferentially detected in Herbig Ae/Fe systems. The residual velocity along the arms does not change sign when crossing the disk minor axis, as confirmed in MWC 758 and HD 135344B, suggesting that the gas in the arms includes a significant non-azimuthal velocity component. 
\item The sub-Keplerian rotation at the outer edge, plausibly arising from the steep negative pressure gradient, is evident in the two-dimensional residual maps for several disks (DM~Tau, HD~135344B, HD~143006, J1604, J1842, LkCa~15, PDS~66, SY~Cha, V4046~Sgr). 
\item Multiple disks show quadrupolar patterns in the velocity residuals, with alternating blueshifted and redshifted residuals across the minor and major axes, which can be caused by geometric deviations in the height of the emitting surface (DM~Tau, J1615, J1852, V4046~Sgr).
\item J1852, PDS~66, and V4046~Sgr appear dynamically more quiescent than the other disks as they show no prominent substructures in the line-width residuals or peak intensity images.
\end{itemize}
Although the sample is biased toward large ($>$1$\arcsec$) disks and is not representative of the full disk population, our results suggest that substructures in CO gas may be relatively common. At the same time, however, the combination of high sensitivity and spatial resolution reveals substantial object-to-object diversity, indicating that detailed, focused investigations of individual sources are well worth pursuing.

\section*{}
\begin{center}
\textbf{Acknowledgment}
\end{center}
This paper makes use of the following ALMA data: ADS/JAO.ALMA\#2021.1.01123.L. ALMA is a partnership of ESO (representing its member states), NSF (USA) and NINS (Japan), together with NRC (Canada), MOST and ASIAA (Taiwan), and KASI (Republic of Korea), in cooperation with the Republic of Chile. The Joint ALMA Observatory is operated by ESO, AUI/NRAO and NAOJ. The National Radio Astronomy Observatory is a facility of the National Science Foundation operated under cooperative agreement by Associated Universities, Inc. We thank the North American ALMA Science Center (NAASC) for their generous support including providing computing facilities and financial support for student attendance at workshops and publications. M.A. acknowledges support from JSPS KAKENHI Grant Number 25K17431. J.B. acknowledges support from NASA XRP grant No. 80NSSC23K1312. M.B., D.F., and J.S. have received funding from the European Research Council (ERC) under the European Union’s Horizon 2020 research and innovation program (PROTOPLANETS, grant agreement No. 101002188). Computations by J.S. have been performed on the "Mesocentre SIGAMM" machine, hosted by Observatoire de la Côte d’Azur. P.C. acknowledges support by the ANID BASAL project FB210003. S.F. is funded by the European Union (ERC, UNVEIL, 101076613), and acknowledges financial contribution from PRIN-MUR 2022YP5ACE. M.F. is supported by a Grant-in-Aid from the Japan Society for the Promotion of Science (KAKENHI: No. JP22H01274). C.H. acknowledges support from NSF AAG grant No. 2407679. J.D.I. acknowledges support from an STFC Ernest Rutherford Fellowship (ST/W004119/1) and a University Academic Fellowship from the University of Leeds. Support for AFI was provided by NASA through the NASA Hubble Fellowship grant No. HST-HF2-51532.001-A awarded by the Space Telescope Science Institute, which is operated by the Association of Universities for Research in Astronomy, Inc., for NASA, under contract NAS5-26555. G.L. has received funding from the European Union’s Horizon 2020 research and innovation program under the Marie Sklodowska-Curie grant agreement No. 823823 (DUSTBUSTERS). C.L. has received funding from the European Union’s Horizon 2020 research and innovation program under the Marie Sklodowska-Curie grant agreement No. 823823 (DUSTBUSTERS) and by the UK Science and Technology research Council (STFC) via the consolidated grant ST/W000997/1. F.Me. has received funding from the European Research Council (ERC) under the European Union's Horizon Europe research and innovation program (grant agreement No. 101053020, project Dust2Planets). C.P. acknowledges Australian Research Council funding via FT170100040, DP18010423, DP220103767, and DP240103290. D.P. acknowledges Australian Research Council funding via DP18010423, DP220103767, and DP240103290. G.R. acknowledges funding from the Fondazione Cariplo, grant No. 2022-1217, and the European Research Council (ERC) under the European Union’s Horizon Europe Research \& Innovation Programme under grant agreement No. 101039651 (DiscEvol). H.-W.Y. acknowledges support from National Science and Technology Council (NSTC) in Taiwan through grant NSTC 113-2112-M-001-035- and from the Academia Sinica Career Development Award (AS-CDA-111-M03). G.W.F. acknowledges support from the European Research Council (ERC) under the European Union Horizon 2020 research and innovation program (grant agreement No. 815559 (MHDiscs)). G.W.F. was granted access to the HPC resources of IDRIS under the allocation A0120402231 made by GENCI. T.C.Y. acknowledges support by Grant-in-Aid for JSPS Fellows JP23KJ1008. Support for B.Z. was provided by The Brinson Foundation. This work was partly supported by the Deutsche Forschungsgemeinschaft (DFG, German Research Foundation)—Ref No. 325594231 FOR 2634/2 TE 1024/2-1, and by the DFG Cluster of Excellence Origins (https://www.origins-cluster.de/). This project has received funding from the European Research Council (ERC) via the ERC Synergy Grant ECOGAL (grant 855130). Views and opinions expressed by ERC-funded scientists are however those of the author(s) only and do not necessarily reflect those of the European Union or the European Research Council. Neither the European Union nor the granting authority can be held responsible for them. Data analysis was in part carried out on the Multi-wavelength Data Analysis System operated by the Astronomy Data Center (ADC), National Astronomical Observatory of Japan. We acknowledge the use of DeepL Translator for improving the English language of this manuscript.

\vspace{5mm}
\facilities{ALMA}

\bibliography{mybib}{}
\bibliographystyle{aasjournal}

\newpage
\appendix
\section{Moment Maps in $^{13}$CO}\label{sec:appendix_13co}
\begin{figure*}
   \centering
   \vspace{1em}   
   \includegraphics[width=\textwidth]{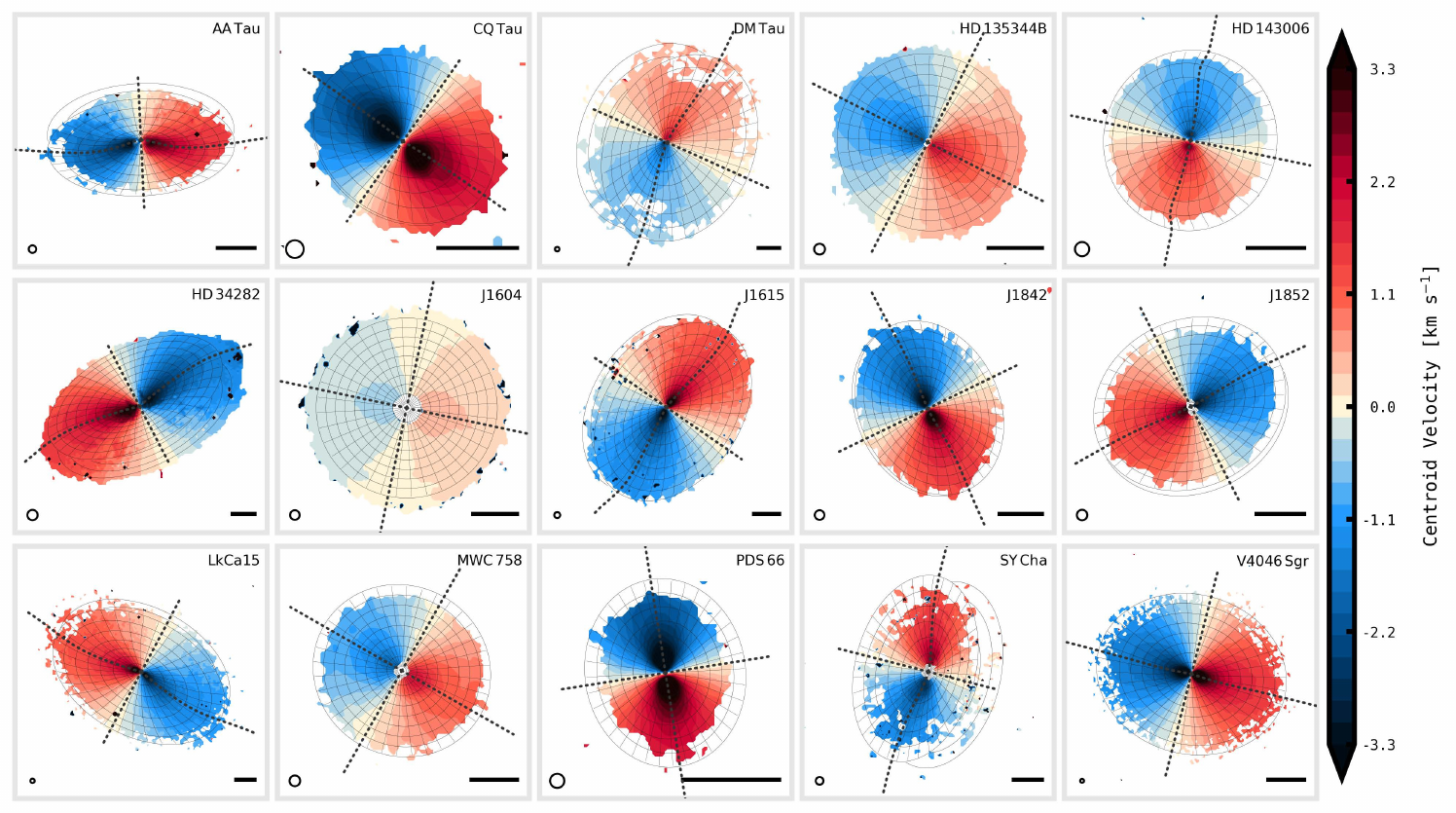}
    \caption{Velocity (line centroid) in $^{13}$CO (3--2). The definition of lines and beam-size indication are the same as Figure~\ref{fig:velocity_12co}. Pixels with S/N below 4.5 are masked.\label{fig:velocity_13co} }
\end{figure*}

\begin{figure*}
   \centering
   \vspace{1em}   
   \includegraphics[width=\textwidth]{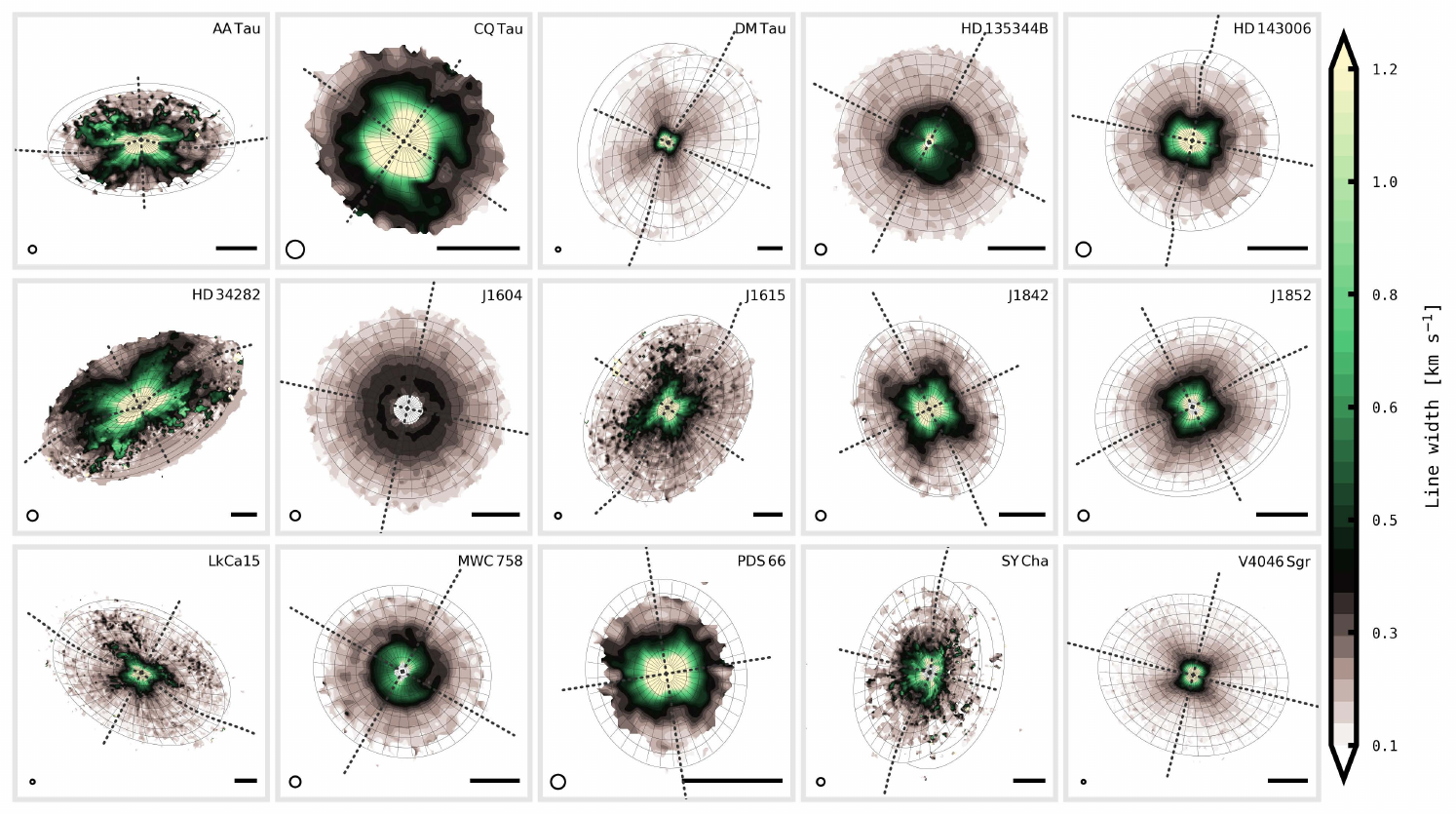}
    \caption{Line width in $^{13}$CO (3--2). Pixels with S/N below 4.5 are masked.\label{fig:linewidth_13co} }
\end{figure*}

\begin{figure*}
   \centering
   \vspace{1em}   
   \includegraphics[width=\textwidth]{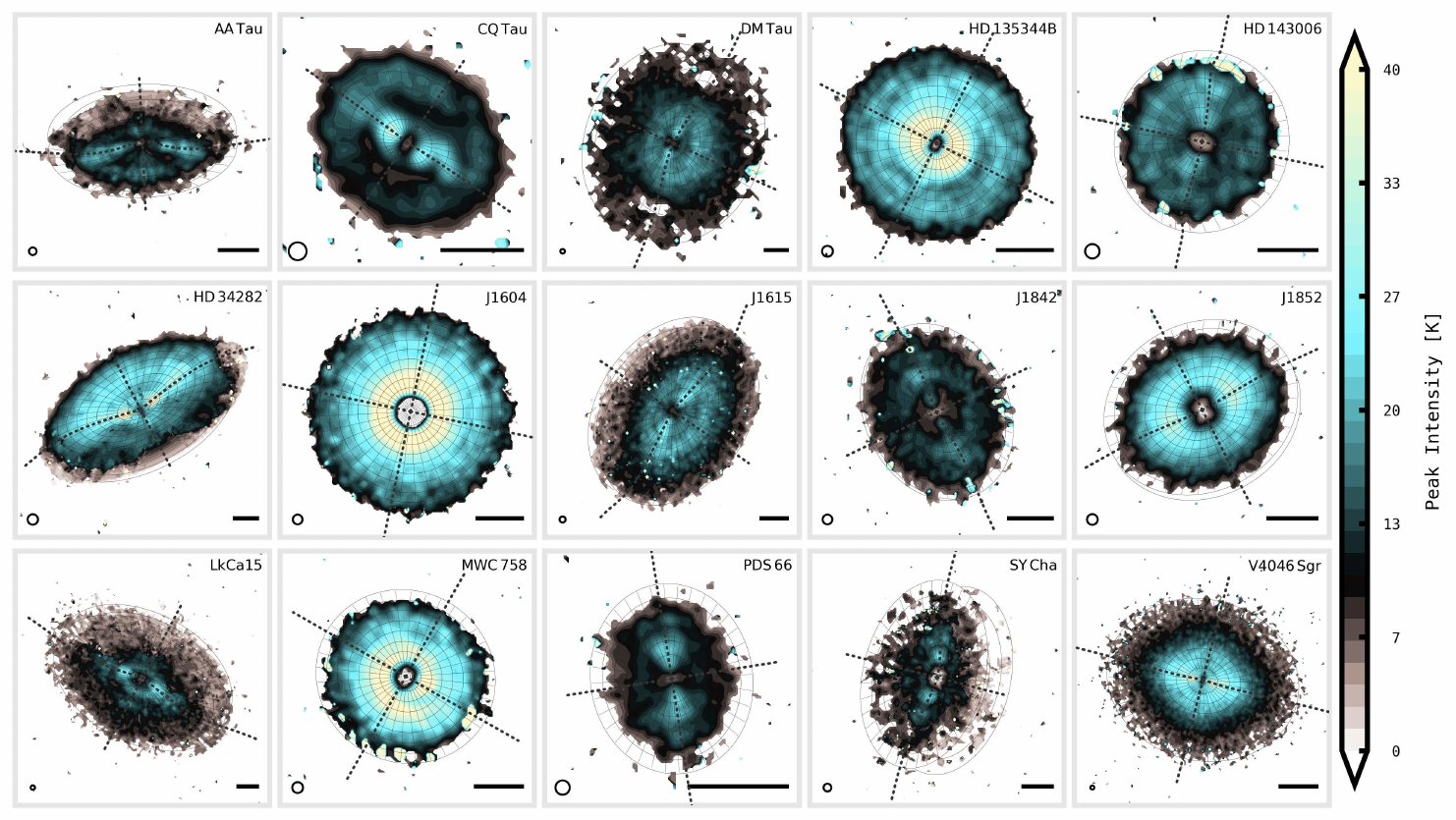}
    \caption{Peak intensity in $^{13}$CO (3--2). Pixels with S/N below 4 are masked.\label{fig:peakint_13co} }
\end{figure*}

\begin{figure*}
   \centering
   \vspace{1em}   
   \includegraphics[width=\textwidth]{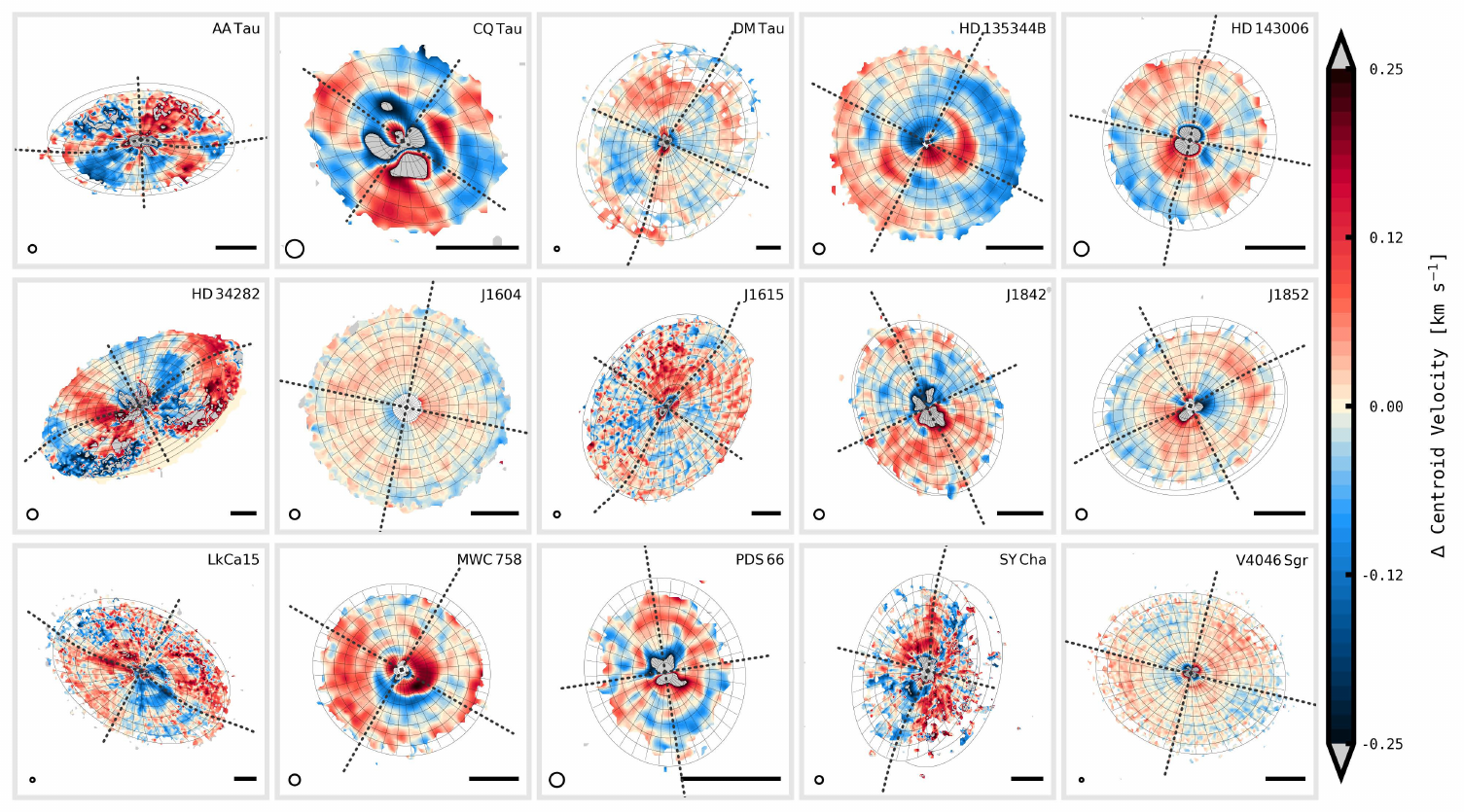}
    \caption{Residual centroid-velocity in $^{13}$CO (3--2). The definition of lines and beam-size indication are the same as Figure~\ref{fig:velocity_12co}. Pixels with S/N below 4.5 are masked.\label{fig:velocity_res_13co} }
\end{figure*}

\begin{figure*}
   \centering
   \vspace{1em}   
   \includegraphics[width=\textwidth]{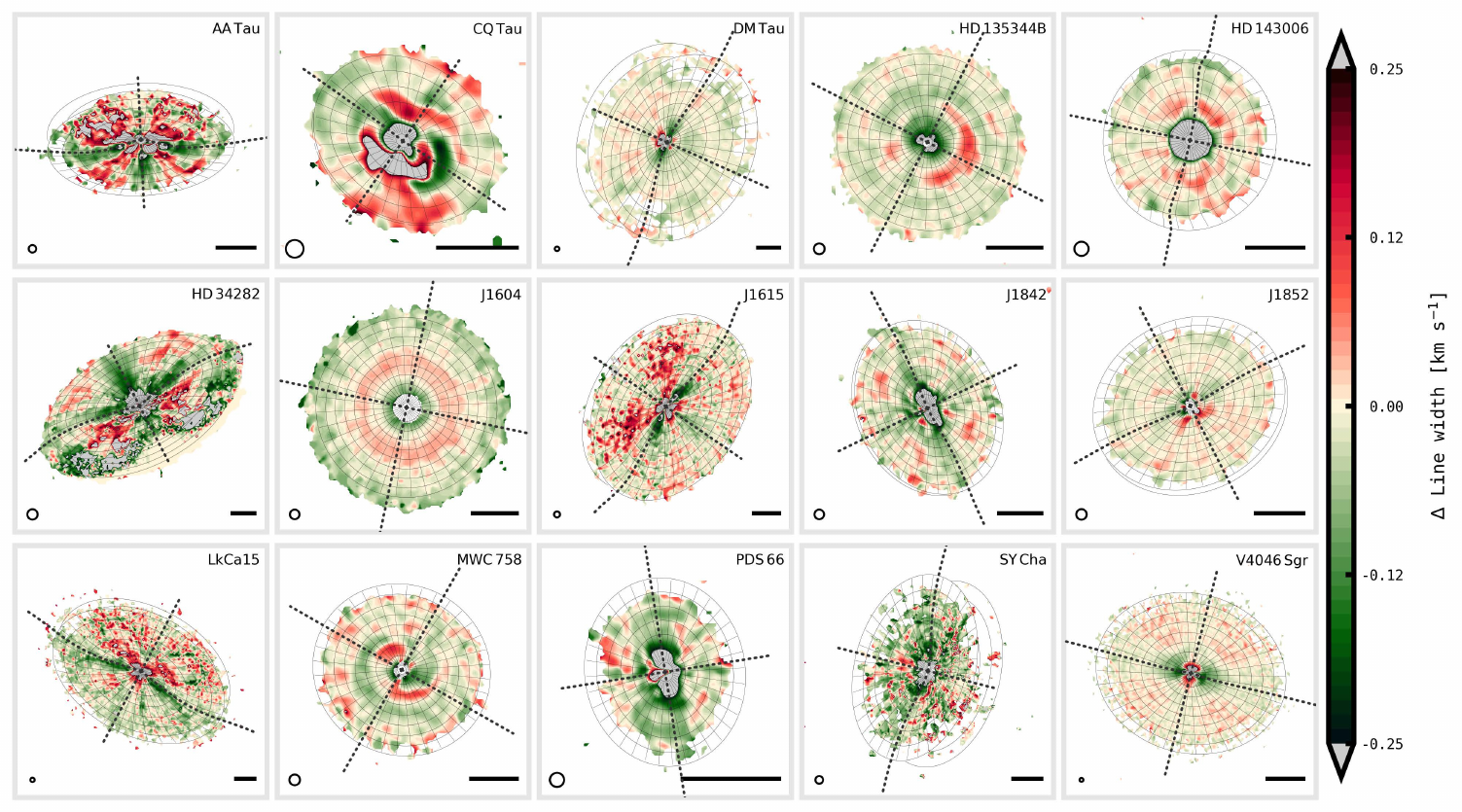}
    \caption{Residual line-width in $^{13}$CO (3--2). Pixels with S/N below 4.5 are masked.\label{fig:linewidth_res_13co} }
\end{figure*}

\begin{figure*}
   \centering
   \vspace{1em}   
   \includegraphics[width=\textwidth]{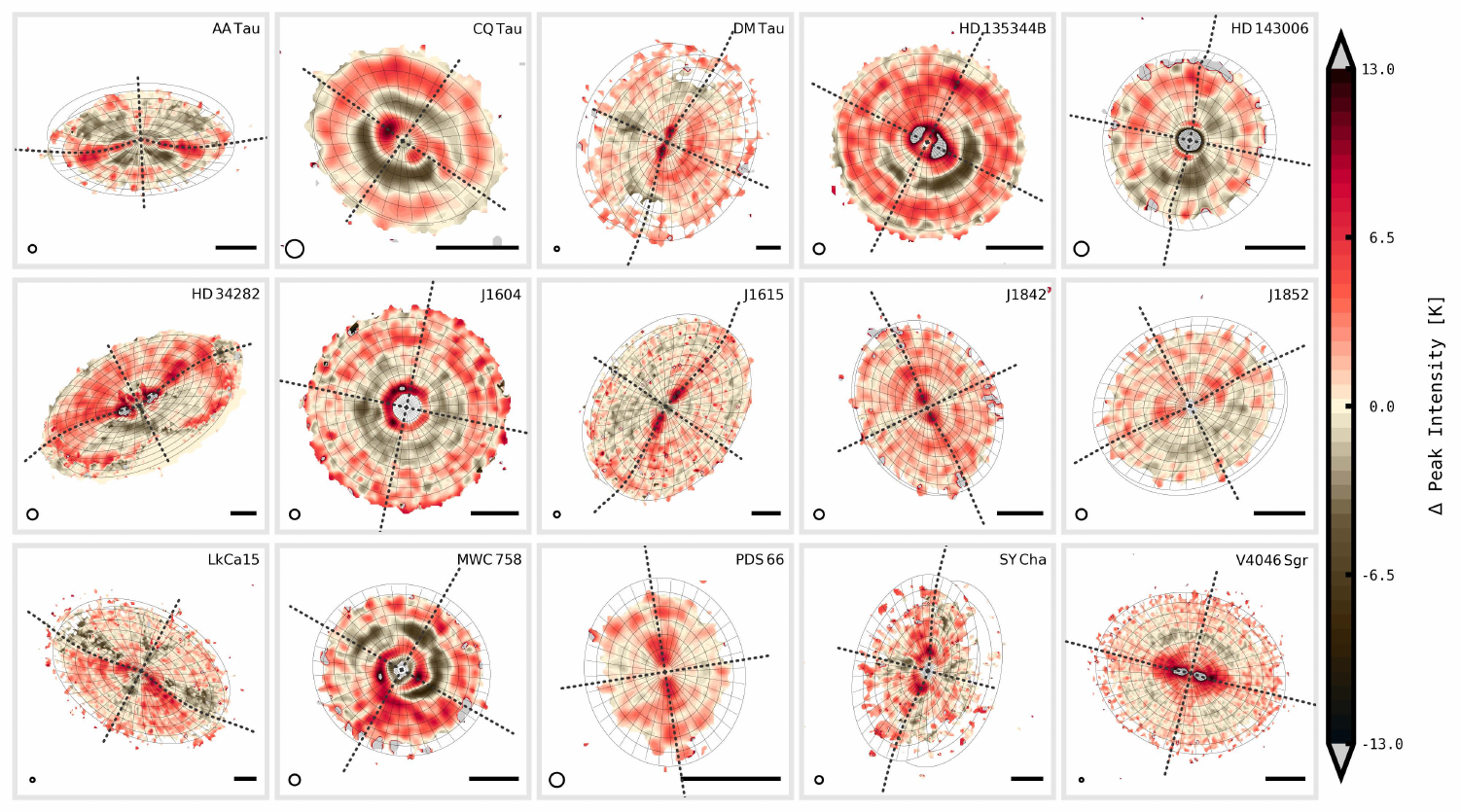}
    \caption{Residual peak intensity in $^{13}$CO (3--2). Pixels with S/N below 4.5 are masked.\label{fig:peakint_res_13co} }
\end{figure*}

The $^{13}$CO images of centroid velocity, line width, peak intensity, and their residuals from the smooth Keplerian disk model are shown in Figures~\ref{fig:velocity_13co}--\ref{fig:peakint_res_13co}. 

\section{Moment Maps in CS}\label{sec:appendix_cs}
\begin{figure*}
   \centering
   \vspace{1em}   
   \includegraphics[width=\textwidth]{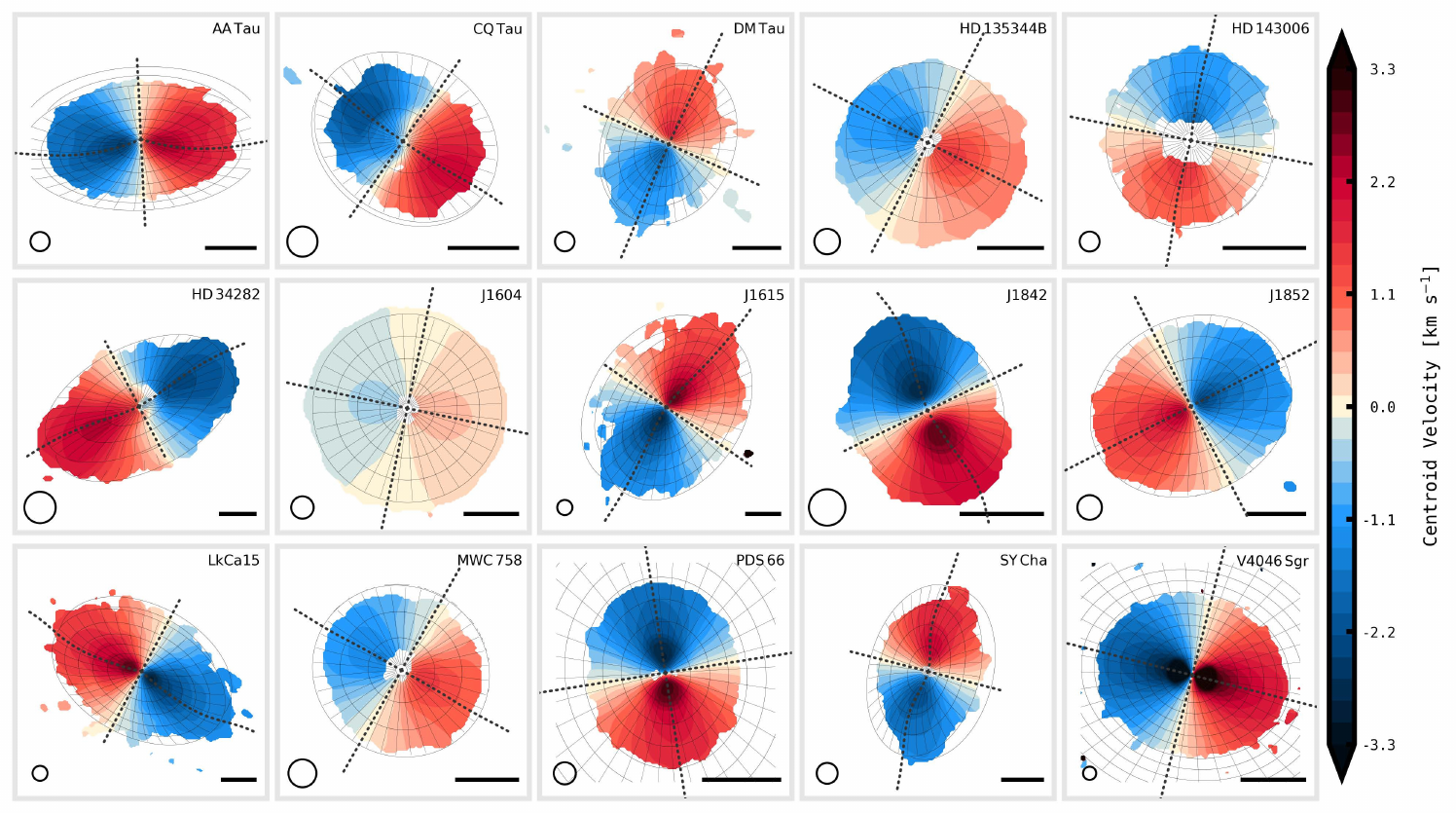}
    \caption{Velocity (line centroid) in CS (7--6). The definition of lines and beam-size indication are the same as Figure~\ref{fig:velocity_12co}. Pixels with S/N below 4.5 are masked. \label{fig:velocity_cs}}
\end{figure*}

\begin{figure*}
   \centering
   \vspace{1em}   
   \includegraphics[width=\textwidth]{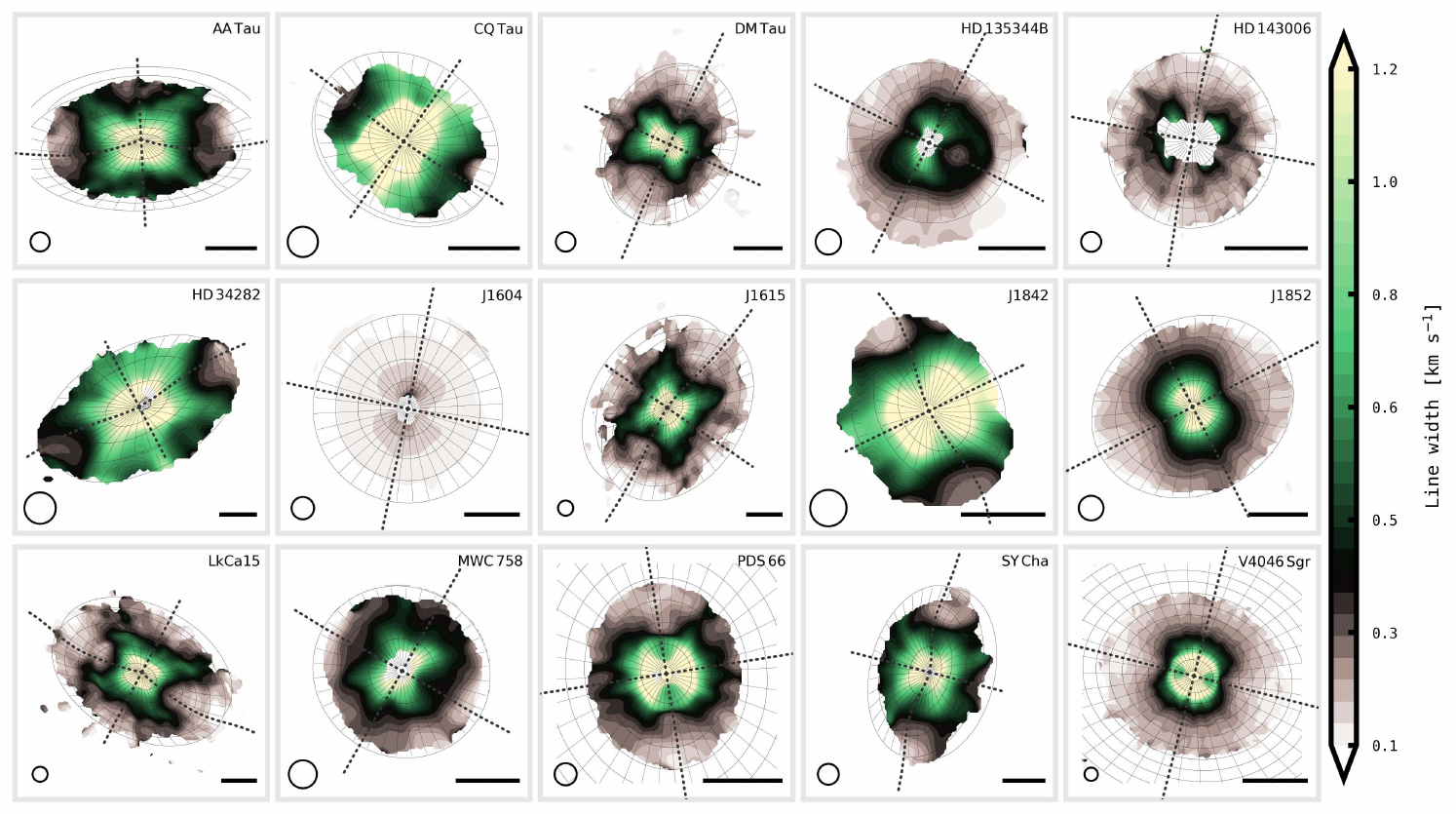}
    \caption{Line width in CS (7--6). Pixels with S/N below below 4.5 are masked. \label{fig:linewidth_cs}}
\end{figure*}

\begin{figure*}
   \centering
   \vspace{1em}   
   \includegraphics[width=\textwidth]{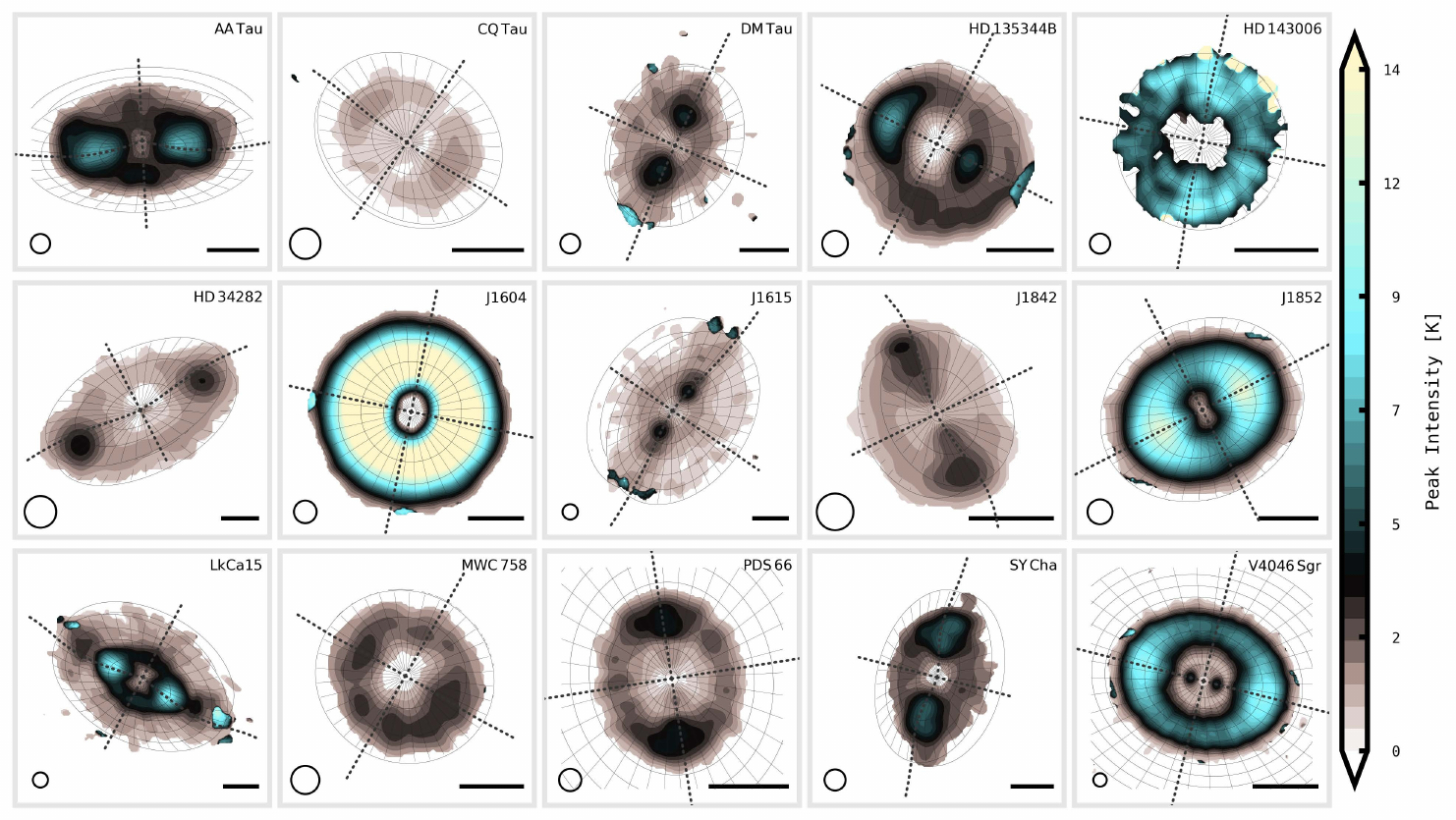}
    \caption{Peak intensity in CS (7--6). Pixels with S/N below 5 are masked.\label{fig:peakint_cs} }
\end{figure*}

\begin{figure*}
   \centering
   \vspace{1em}   
   \includegraphics[width=\textwidth]{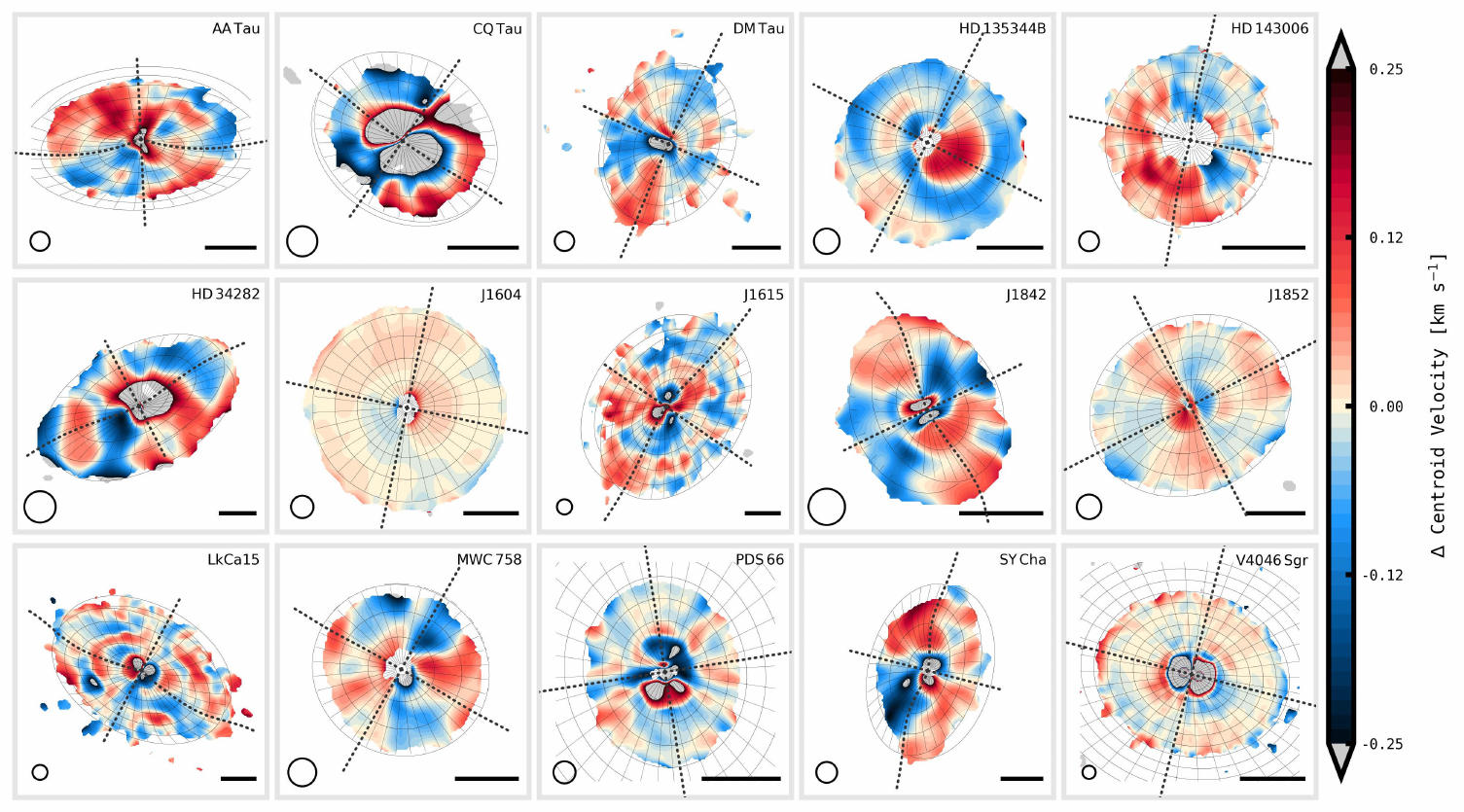}
    \caption{Residual centroid-velocity in CS (7--6). The definition of lines and beam-size indication are the same as Figure~\ref{fig:velocity_12co}. Pixels with S/N below 4.5 are masked. \label{fig:velocity_res_cs}}
\end{figure*}

\begin{figure*}
   \centering
   \vspace{1em}   
   \includegraphics[width=\textwidth]{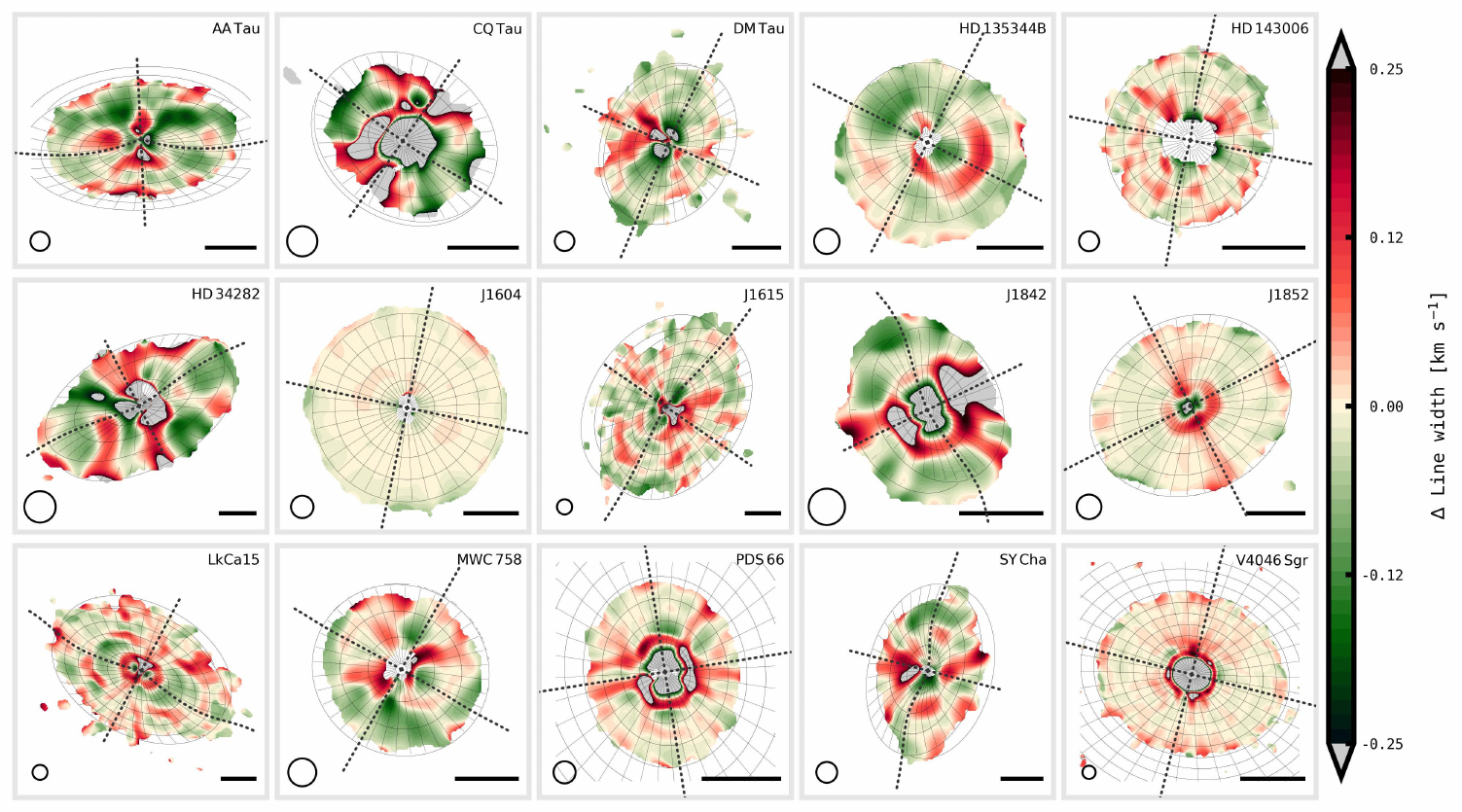}
    \caption{Residual line-width in CS (7--6). Pixels with S/N below 4.5 are masked.\label{fig:linewidthy_res_cs}}
\end{figure*}

\begin{figure*}
   \centering
   \vspace{1em}   
   \includegraphics[width=\textwidth]{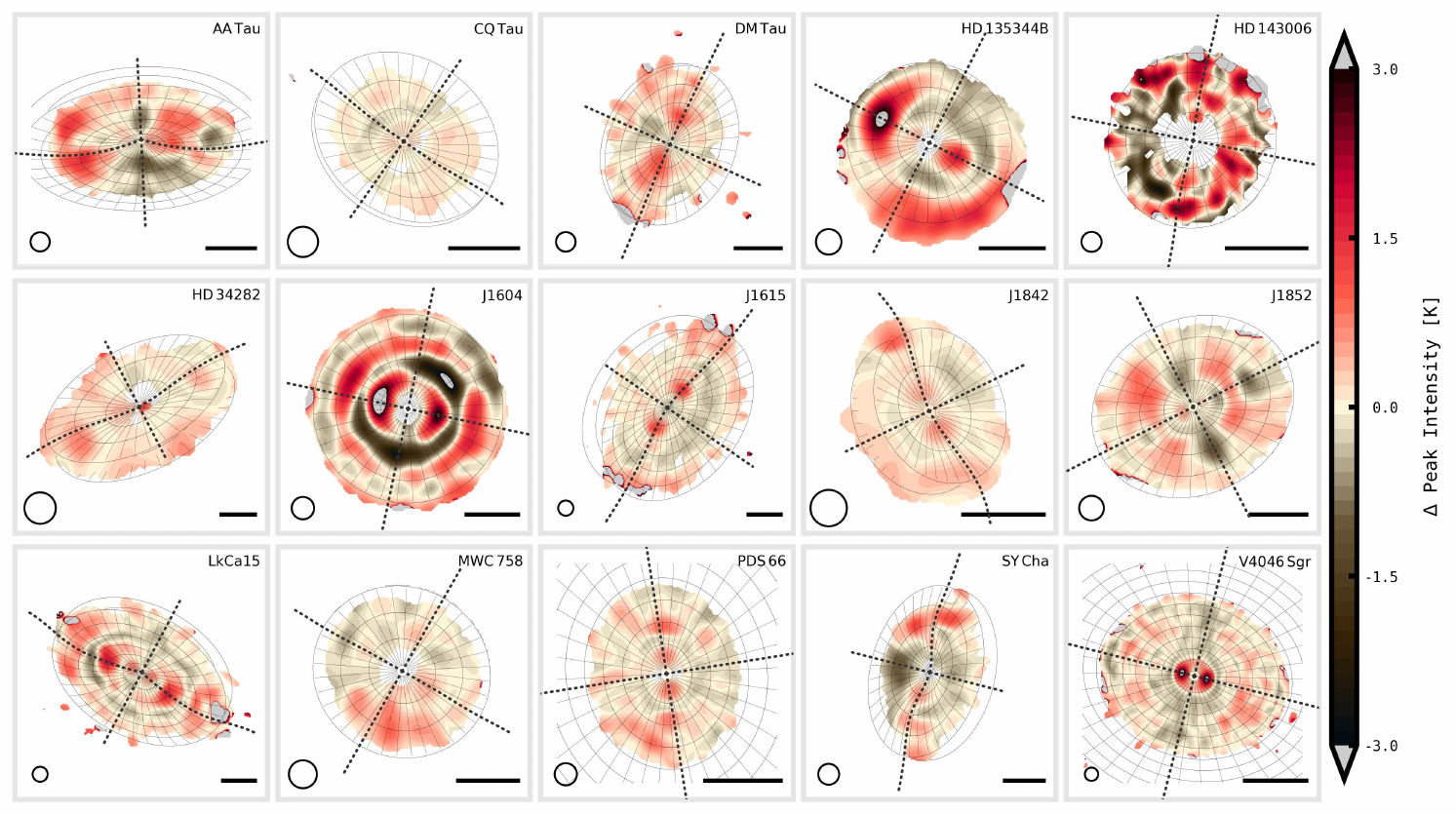}
    \caption{Residual peak-intensity in CS (7--6). Pixels with S/N below 5 are masked.}\label{fig:peakint_res_cs} 
\end{figure*}

Because CS (7--6) emission is fainter than $^{12}$CO (3--2) and $^{13}$CO (3--2), the CS images are produced with a coarse angular and velocity resolutions of $0\farcs3$ and 200 m\,s$^{-1}$, respectively, for most sources (Table~\ref{tab:beam_chan}). The images of centroid velocity, line width, peak intensity, and their residuals from the smooth Keplerian disk model are shown in Figures~\ref{fig:velocity_cs}--\ref{fig:peakint_res_cs}. 

\begin{deluxetable}{lllllll}
\tabletypesize{\scriptsize}
\tablecaption{Beam size and channel spacing for the images in this paper}\label{tab:beam_chan}
\tablewidth{0pt}
\tablehead{
\colhead{Source} & \multicolumn{2}{c}{$^{12}$CO}  & \multicolumn{2}{c}{$^{13}$CO} & \multicolumn{2}{c}{CS}\\
 & beam & chan.  & beam & chan.  & beam & chan. \\
  & ($\arcsec$) & (km s$^{-1}$) &  ($\arcsec$) &  (km s$^{-1}$) &  ($\arcsec$) &  (km s$^{-1}$)
}
\startdata 
AA Tau&0.15&0.1&0.15&0.1&0.3&0.2 \\
CQ Tau&0.15&0.1&0.15&0.1&0.3&0.2\\
DM Tau&0.15&0.1&0.15&0.1&0.3&0.2\\
HD 135344B&0.15&0.1&0.15&0.1&0.3&0.2\\
HD 143006&0.15&0.1&0.15&0.2&0.15&0.2\\
HD 34282&0.15&0.1&0.15&0.1&0.3&0.2\\
J1604&0.15&0.028&0.15&0.1&0.3&0.1\\
J1615&0.15&0.1&0.15&0.2&0.3&0.2\\
J1842&0.15&0.1&0.15&0.2&0.3&0.2\\
J1852&0.15&0.1&0.15&0.1&0.3&0.2\\
LkCa 15&0.15&0.1&0.15&0.1&0.3&0.2\\
MWC 758&0.15&0.1&0.15&0.2&0.3&0.2\\
PDS 66&0.15&0.028&0.15&0.2&0.3&0.2\\
SY Cha&0.15&0.1&0.15&0.2&0.3&0.2\\
V4046 Sgr&0.15&0.1&0.15&0.1&0.3&0.2\\
\enddata
\tablenotetext{}{Details of the \textsc{discminer} fitting, including the adopted kernel and the best-fit parameters, can be found in \cite{exoALMA_discminer}.}
\end{deluxetable}

\end{document}